# Polarized and narrow excitonic emission from graphene-capped monolayer $WS_2$ through resonant phonon relaxation

Garima Gupta and Kausik Majumdar*

*Department of Electrical Communication Engineering, Indian Institute of Science, Bangalore 560012, India*

**Corresponding author, E-mail: kausikm@iisc.ac.in*

**Abstract:** The broadening and polarization of excitonic luminescence in monolayer TMDs largely suffer from inhomogeneity and temperature – an unresolved problem to date. In this work, through few-layer-graphene encapsulation of monolayer $WS_2$, we reduce the inter-excitonic energy separation, which then can have a narrow resonance with a specific phonon mode of our choice. The resulting single-step exciton relaxation with the resonating phonon mode significantly suppresses the inhomogeneous broadening, allowing us to achieve the narrowest exciton linewidth of 1.06 meV (which translates to 0.19 meV after deconvolution with the excitation laser linewidth). The single-phonon resonance helps to achieve a high quantum efficiency despite graphene encapsulation. The technique is powerful in tuning the exciton polarization during relaxation by choosing a specific resonating phonon mode. For example, the valley coherence (polarization) improves from $\sim$68% ($\sim$40%) to $\sim$90% ($\sim$75%) on resonance with $2A_1^{'}$ and $A_1'$ modes respectively. We further demonstrate a strong polarization reversal on resonance with a chiral phonon mode. Strikingly, the above features remain robust against temperature (up to 200 K) and sample age (few months in ambient condition). The findings will lead to clean excitonic measurements without requiring cryogenic cooling.

**Introduction:** Despite being ultrathin, the monolayers of Transition Metal Dichalcogenides (TMDs) are highly luminescent because of the strongly bound 2D excitons [1–4]. However, it is practically impossible to avoid the influence of the surrounding environment on the monolayers [5], obscuring the innate excitonic features [6]. The presence of potential fluctuations from external (adsorbed or substrate charge impurities) [7] and internal (defects) [8–11] factors leads to a large inhomogeneous broadening of the photoluminescence (PL) peaks [12]. Several methods, such as chemical treatment of the substrate [13], and encapsulation with hBN [14–17] or graphene [18] flakes have been explored to reduce the inhomogeneous broadening. Encapsulation with graphene not just screens the substrate induced inhomogeneity, but also suppresses emission from other excitonic species such as defect bound excitons, trions, charged biexcitons [18]. Additionally, it also helps in retaining a high valley coherence and valley polarization of the excitons due to suppression of the electron-hole exchange interaction [19]. However, one drawback of this technique is reduced quantum efficiency and hence weak peak intensity due to fast carrier transfer to graphene during the relaxation process. In addition, the applicability of all these techniques is largely limited to cryogenic temperatures only. The room temperature PL inevitably suffers from broad linewidth and negligible valley polarization. No definite solution is reported to date which can provide narrow-linewidth, highly polarized exciton peaks at room temperature.

In this work, we overcome these issues by using a simple technique: (a) We sandwich a monolayer $WS_2$ between two few-layer-graphene (FLG) flakes such that the resulting screening makes the inter-excitonic energy separation $(\Delta_{ns-1s}; n > 1)$ approximately equal to an optical phonon energy in the system. (b) Using a tunable laser, we resonantly generate the higher energy excitons $(ns)$, which undergo ultrafast relaxation to the lowest energy $1s$ state in a single step, by scattering with the resonating phonon mode. As an outcome of this technique, we achieve, (1) high intensity, (2) ultra-narrow linewidth (1.06 meV) exciton peak which also shows (3) a high valley coherence measured by the degree of linear polarization (DOLP $\sim$ 90 %) and valley polarization measured by the degree of circular polarization (DOCP $\sim$ 75 %). These features remain robust till $200$ K, making this technique appealing for several experiments and applications.

**Various mechanisms of relaxation**: To understand the working principle of this technique, let us revisit the process of exciton generation, relaxation, and light emission from radiative recombination. Initially, the higher energy excitons inside the light cone of the $ns$ band, that resonate with the incoming photon energy $(hc/\lambda_{ex})$ are generated (Fig. 1). These excitons can relax to the $1s$ band through a combination of the following mechanisms: M1 - losing energy through multiple scatterings with various phonon modes in the system (Fig. 1a), M2 - through direct optical transition by photon emission between the dipole coupled Rydberg states (Fig. 1b), and M3 - as explored in this paper as a special case of M1 - by scattering with a single resonating zone-center optical phonon mode and relaxing to the $1s$ state in one step (Fig. 1c). Finally, the excitons inside the light cone of the $1s$ band recombines radiatively to emit light of energy $hc/\lambda_{em}$. Due to both incoming and outgoing resonance, we expect this single-step scattering to dominate the $ns \rightarrow 1s$ transition. The implications of this single-step relaxation on emission are discussed in the upcoming sections.

**Experimental Observations**: The emission spectrum of the FLG-$WS_2$-FLG (GWG) stack (Fig. 2a) (see Methods for sample preparation) taken with 532 nm laser excitation at 5 K shows conspicuous 1s and 2s exciton peaks (Fig. 2b), with the 1s peak exhibiting an average linewidth of ~9.5 meV (inset). Due to the FLG induced screening, we obtain $\Delta_{2s-1s} \sim 31.5$ meV, which matches well with the previously reported value [20] and implies a significant reduction in the exciton binding energy. The large red shift of the higher energy exciton states [21] brings the inter-excitonic separation in the regime of optical phonon energies.

We carry out the excitation (PLE) spectroscopy of this GWG stack and monitor emission from the $1s$ state. The excitation wavelength is tuned from 570 to 600 nm using a supercontinuum laser source and a monochromator setup (see Methods for measurement details). Fig. 2c (left panel) shows the color plot of the $1s$ intensity (in log scale) as a function of $\lambda_{ex}$ and $\lambda_{em}$. The corresponding intensity color plot (in linear scale), normalized at each $\lambda_{ex}$, is shown in the right panel. At specific values of $\lambda_{ex}$, where the resonance condition $\Delta_{ns\rightarrow 1s} \approx \hbar\omega_{ph}$ is satisfied, we observe highly luminescent and narrow linewidth exciton peaks through mechanism M3 (see Supplemental Material [22] for individual spectrum). The resonating phonon modes are labelled in Fig. 2c [23,24]. The line cuts show the individual spectrum in Fig. 2d-2f. The broad pink backgrounds represent emission from excitons formed through multi-phonon relaxation process (M1). The sharp green peaks in Fig. 2d and 2e correspond to excitons formed through single-phonon (M3) and two-phonon resonance, respectively. The exciton peak at single- (two-) phonon resonance is remarkably narrow and about 124-fold (3.4-fold) enhanced (Fig. 2d-e), as compared to the broad and weakly luminescent exciton peak in Fig. 2f, when such resonance is absent.

**Reduction in the inhomogeneous linewidth**: External and internal inhomogeneities cause potential fluctuations in the monolayer. The resulting random doping and stark effect lead to a spatial perturbation in the exciton peak position (Fig. 3a). This leads to a real space to energy mapping of the excitons. The intersections of the solid lines with the spatial exciton energy distribution in the top panel of Fig. 3a are the equi-energy spots corresponding to a specific exciton energy in the real space. The bottom inset of Fig. 3b schematically shows such equi-energy locations under the laser spot at different $1s$ exciton energies. The inhomogeneous broadening is a result of collecting photons from all such spatially distributed $1s$ states under the laser spot (Fig. 3b and left panel of Fig. 3c).

The proposed technique relies on exploiting the narrow spectral resonance between the incoming excitation and a specific $1s$ exciton energy state via scattering with the resonating phonon mode (Fig. 3b and right panel of Fig. 3c). To explain this, let us look closely into the case when $A_1'$ (418 $cm^{-1}$) is the resonating phonon mode in the relaxation process for $589 \leq \lambda_{ex} \leq 600$ nm. As estimated from the Bethe-Salpeter equation, $\lambda_{ex}$ in this range generates excitons in the bands ranging from $6s \leq ns \leq 3s$ (see Note S1 in Supplemental Material [22]). On the excitation side, only the energy states in the inhomogeneously broadened $ns$ exciton band, that resonate in energy with $hc/\lambda_{ex}$ are excited (green laser spikes, Fig. 3b). Thus, the incoming laser performs a preliminary spectral (and hence spatial) filtering by generating excitons at the $ns$ state at specific locations in the real space. Further, as $\Delta_{ns-1s}$ ($\approx 52$ meV) is around the energy of the $A_1'$ phonon mode in $WS_2$, relaxation to the $1s$ state is dominated by scattering with the $A_1'$ phonons in a single step through mechanism M3 (dashed arrows, Fig. 3b). As this relaxation process is ultrafast, only those $1s$ energy states in the inhomogeneously broadened peak, that lie at an energy $\hbar\omega_{A_{1g}}$ below $hc/\lambda_{ex}$, are populated (solid red spikes, Fig. 3b). This leaves little chance for populating the other states in the distribution through slow, multi-phonon relaxation mechanisms (such as M1), truncating the inhomogeneous broadening significantly. This implies that we are essentially collecting light only from the equi-energy spots in the real space that correspond to those resonating $1s$ states.

The experimental demonstration of the above mechanism showing light emission from the specifically populated resonant $1s$ states is shown in Fig. 3d. As we tune $\lambda_{ex}$, different regions of the inhomogeneously distributed $1s$ states satisfying $\Delta_{ns-1s} \approx \hbar\omega_{A_1'}$ are populated and emit light (shaded orange peaks). For comparison, we also plot the inhomogeneously broadened envelope obtained at $\lambda_{ex} = 532$ nm (in grey), which is a collective outcome of emission from all these individual peaks in the distribution (through M1). This shows the supremacy of our technique, which makes excitonic emission much less sensitive to the surrounding inhomogeneity. This is also evident from the invariance of the emission linewidth of individual peaks (see Fig. S4 in Supplemental Material [22]). The data from another GWG sample is shown in Fig. S5 of Supplemental Material [22].

To support the argument, we perform the emission imaging with 591, 593 and 595 nm excitation wavelengths in Fig. 3e. We observe a significant modulation in the spatial distribution of the luminescence peaks (marked by arrows in Fig. 3d), suggesting varying spatial profiles of the different equi-energy spots.

We obtain an average total linewidth of $1.21 \pm 0.04$ meV (Fig. 3f for a representative plot), with the lowest linewidth being 1.06 meV (see Fig. S6 in Supplemental Material [22]). These numbers are lower than the narrowest linewidth reported to date from free excitons in TMDC monolayers [13,14,16,25,26]. Please note that the difference between the laser linewidth ($\sim 0.8$ meV) and the narrowest obtained exciton linewidth is well above our system resolution of $\sim 40 - 60$ μeV. On deconvoluting with the excitation laser having nonzero broadening, we extract an exciton linewidth of $\sim 0.19 \pm 0.07$ meV, which denotes an upper bound on the homogeneous broadening of the $1s$ exciton states in GWG stack. This implies a net exciton lifetime $\geq 1.65$ ps in the system. This timescale is in good agreement with the reported lifetime values obtained from time-resolved PL measurements [18,27]. This is the first report of achieving an excitonic emission linewidth down to the true homogeneous limit. Due to suppressed inhomogeneity, we also observe a tighter distribution of the linewidth with the standard deviation to mean ratio reducing from 8% under off-resonant excitation to 3% with this technique.

**Enhanced Quantum efficiency in spite FLG encapsulation**: The internal quantum efficiency of emission in the proposed mechanism can be written as a product of relaxation ($\eta_{rel}$) and emission ($\eta_{em}$) quantum efficiency:

$$\eta_{int} = \underbrace{\frac{\gamma_{res}}{\gamma_{res} + \gamma_{off-res} + \gamma_{gr}}}_{\eta_{rel}} \times \underbrace{\frac{\gamma_r}{\gamma_r + \gamma_{nr} + \gamma_{gr}}}_{\eta_{em}} \tag{1}$$

As a result of the ultrafast relaxation rate ($\gamma_{res}$) in M3: (1) The generated oscillator strength is primarily transferred to the resonating final $1s$ states. This avoids the distribution of the oscillator strength among other inhomogeneously distributed states, like in the case of other relaxation processes such as M1 ($\gamma_{off-res}$). (2) The loss due to exciton transfer to the top and bottom FLG ($\gamma_{gr}$) during such fast relaxation is also minimized. Therefore, as $\gamma_{res} \gg \gamma_{off-res}$ and $\gamma_{gr}$, the generated oscillator strength efficiently relaxes from the $ns$ band to $1s$ band due to dual resonance, thus making $\eta_{rel}$ large. This is evident in Fig. 2d by the large ratio of the intensity of the green curve (M3) to the pink background (M1). $\gamma_r$ and $\gamma_{nr}$ represent the exciton radiative and non-radiative (other than exciton transfer to FLG) decay rates once it relaxes to the 1s band. For simplicity, we assumed that $\gamma_{gr}$ is energy independent in equation 1.

**High exciton valley polarization and valley coherence**: Excitons inherit the angular momentum of the electronic bands in the $K, K'$ valley. Therefore, an incoming circularly (linearly) polarized light generates exciton in a specific valley (superposition state of both the valleys) [28–30]. During (1) the relaxation process, and (2) inside the light cone, the degradation in the exciton valley polarization/coherence occurs due to a coupled effect of exciton scattering and pseudospin presession due to exchange interaction [19,31] (Fig. 4a). In addition, during the relaxation process M1, the scattering with multiple phonons with different polarizations may also lead to a random phase accumulation. We observe these effects by the gradual improvement in the exciton DOLP, for the broad background peak (in pink) in Fig. 2d-f, as the incoming excitation is tuned closer to the $1s$ resonance (open red circles in Fig. 4b).

The FLG encapsulation in the GWG sample helps to suppress both these effects. First, in mechanism M3, as we selectively choose the phonon mode that dominates the single-step relaxation, the uncontrolled pseudospin randomization during the relaxation process is avoided. For example, the polarization of the incoming light does not change by scattering with the $A_1'$ phonon mode due to the diagonal form of its Raman tensor [32,33] (schematically depicted in the inset of Fig. 4d-e). This is depicted by the fully polarized $A_1'$ Raman peak in Fig. 4c [top (bottom) panel for linear (circular) polarization] as obtained using 532 nm excitation. Hence, when we achieve a resonance with the $A_1'$ mode in the exciton relaxation process, in Fig. 4b, we observe a negligible change in the exciton DOLP (red solid symbols) and DOCP (blue solid symbols) on varying $\lambda_{ex}$. Second, once the exciton relaxes down to the $1s$ light cone, the FLG encapsulation helps to preserve the polarization inside the light cone by (a) reducing pseudospin precession frequency due to screening induced suppression of the exchange interaction, and (b) reducing the exciton lifetime due to fast charge transfer to FLG [19]. As a combined effect, we observe a DOLP of $\sim 90$ % and a DOCP of $\sim 75$ % of $1s$ excitons in the GWG stack (Fig. 4d-e).

On the other hand, when the $2A_1^{'}$ phonon mode dominates the relaxation process, we obtain a degraded DOLP (DOCP) of $\sim 68.3$ % ($\sim 40\%$) (Fig. 4f-g). This suggests that the additional exchange interaction in the intermediate state between the two consecutive scattering events during the relaxation results in the observed polarization degradation, as schematically shown in the inset of Fig. 4f-g. This technique can also tune the exciton polarization strategically during the relaxation process if we choose the appropriate phonon mode. For example, during the exciton-phonon scattering, complete reversal of the incoming polarization is possible on resonating with a chiral phonon mode, to ensure angular momentum conservation in $C_{3h}$ symmetric system (schematically depicted in the inset of Fig. 4h). We observe such reversal using the chiral 'b' phonon mode [34–37] at $420$ cm$^{-1}$ in a FLG-$MoS_2$-FLG sample, upon resonating with which, the outgoing photon is reverse circularly polarized with respect to the incoming circular polarization (Fig. 4h).

**Temperature robust exciton linewidth and polarization**: When sample temperature increases, upon relaxation through mechanism M1, the states outside the light cone get populated and start contributing to light emission mediated by phonons (Fig. 5a). These momentum-dark states are long-lived, which give sufficient time to the exciton population to achieve thermal equilibrium before radiative recombination. This enhances the homogeneous linewidth with $k_BT$ [38,39].

In the current technique, direct light cone to light cone transition happens from the higher energy band to the $1s$ band. Therefore, the excitons directly relax to the lowest energy states in the $1s$ band at all temperatures. As $\gamma_r$ and $\gamma_{gr}$ ($\gg \gamma_{nr}$) are independent of temperature, the overall exciton lifetime ($\tau \sim \frac{1}{\gamma_r+\gamma_{gr}}$) inside the light cone remains temperature independent in this process. Since phonon absorption process ($\gamma_{ph,Abs}$) is much slower than the net exciton decay rate (i.e., $\gamma_{ph,Abs} \ll \gamma_r + \gamma_{gr}$), the excitons can hardly move outside the light cone. Consequently, the excitons do not have enough time to attain thermal equilibrium before radiative recombination (Fig. 5b). Therefore, because of this non-equilibrium exciton population inside the light cone, we observe a relatively weak influence of temperature on the linewidth through this technique (Fig. 5c-d) – a key aspect of this work.

Exciton polarization degrades with temperature when it relaxes by mechanism M1, because of two primary reasons: (1) *Enhancement in the overall phonon scattering rate* - As more phonon modes start participating in the exciton relaxation process, the random phase accumulation during relaxation increases with temperature; and (2) *Enhanced contribution of excitons outside the light cone* – As explained previously, the fractional contribution of excitons outside light cone increases with temperature due to relaxation mechanism M1 (Fig. 5a). This in turn has two effects on polarization degradation. First, the net exchange interaction ($J$) increases with the exciton center of mass momentum ($\mathbf{Q}$) as $J \propto -V(\mathbf{Q})|\mathbf{Q}|^2$ where $V(\mathbf{Q})$ is the electron-hole Coulomb interaction [40]. This leads to a faster pseudospin dephasing for excitons with larger $|\mathbf{Q}|$. Second, the excitons outside the light cone being momentum dark, are long-lived, and hence suffer from more accumulation of random phase in the pseudospin vector. In the proposed technique, by minimizing all these processes through relaxation process M3, we observe a dramatically negligible influence of temperature on the exciton valley coherence (Fig. 5c-d).

**Discussion:** Apart from 1s emission, the resonant Raman scattered photons also energetically coincide with the emitted light, and often it becomes challenging to segregate these two mechanisms [41,42]. While the emission process is a result of a cascade of photon absorption, relaxation through exciton-phonon scattering, and photon emission through recombination of the 1s exciton, the resonant Raman process is an instantaneous process maintaining coherence. We argue that the 1s excitonic emission has the dominant contribution over resonant Raman scattering in the current scenario of dual resonance based on the following observations: (a) We note a significantly lower than 100% DOCP and DOLP in the light output (Fig. 4d-g), while the pure Raman process should yield 100% degree of polarization (see, for example, Fig. 4c) for $A_1'$ and $2A_1'$ modes possessing diagonal Raman tensor. (b) The light output has DOCP < DOLP in Fig. 4d-g, which is a characteristic of 2D exciton emission due to the in-plane nature of the pseudo-magnetic field generated by the exchange interaction. (c) The strong correlation between the spectral feature of the non-resonant 1s exciton emission and the envelope of the output under dual resonance when the excitation wavelength is tuned (Fig. 3d) suggests the key role played by the 1s state in the process, which should not happen for Raman scattering. (d) Strikingly, when the emission quantum efficiency is suppressed by increasing the number of layers of $WS_2$, the output photon count becomes independent of the spectral feature of the 1s state, indicating a Raman scattering dominated process for few-layers (details in Note S2 in Supplemental Material [22]). (e) We also observe a dramatic reduction in the peak intensity in the few-layer $WS_2$ samples. Being an instantaneous process, if resonant Raman scattering would have dominated the peak in the monolayer, we would expect similar intensity in few-layer samples as well, suggesting the peak intensity is dominated by excitonic emission in monolayer. (f) The noticeable linewidth difference between the excitation laser and the emission peak (Fig. 3f) is not observed for the $A_1'$ Raman peak obtained with off-resonant excitation in the GWG stack, in which case the Raman linewidth closely matches the linewidth of the excitation laser (see Note S2 in Supplemental Material [22]).

Due to protective encapsulation by FLG and the intrinsic nature of narrow resonance, we find the emission features from the sample to remain robust for a long time (see representative data taken on the same stack after two months in Fig. S7 in Supplemental Material [22]). The strength of the proposed technique lies in obtaining the narrowest excitonic peak reported to date even in the presence of sample inhomogeneity – allowing us to achieve close to the true homogeneous linewidth in the sample. This can help experiments (including Stark effect, Zeeman splitting, Rabi splitting) where the observation of small excitonic peak shift is usually obscured due to large linewidth. The sharpness of the peak along with the observed high degree of exciton polarization is robust even at higher temperatures – making the technique promising for several experiments which earlier required cryogenic temperature to achieve these features. The combination of enhanced quantum efficiency, narrow linewidth, high degree of valley polarization and coherence, and the robustness of these features against sample age and temperature makes the proposed technique powerful to obtain a near-ideal atomically thin light emitter.

## Methods

**Sample preparation:** To prepare the GWG stack, the material is first exfoliated on a Polydimethylsiloxane (PDMS) sheet. After locating an appropriate flake, it is transferred on a Si substrate covered with 285 nm thick thermally grown $SiO_2$, one by one. The thickness of the bottom and the top graphene is 2-3 nm. The sample is annealed at $200^0$C for 5 hours (pressure ~$10^{-6}$ torr) to ensure good adhesion between successive layers.

**Measurement details:** All the measurements are taken in a closed-cycle optical cryostat (Montana Instruments) at 5 K using a $\times 50$ long-working-distance objective having a numerical aperture of 0.5. The photoluminescence excitation measurement is carried out using a supercontinuum laser source (NKT Photonics). The output of the laser is passed through an acousto-optic tunable filter (AOTF) followed by one or two monochromators (Edmund Optics) to ensure spectrally narrow excitation pulses. In all the wavelength dependent measurements, the excitation wavelength is tuned by a step of 1 nm. All measurements except the narrowest emission linewidth measurements were taken with one monochromator. The laser linewidth with one monochromator is ~1.2 meV. The narrowest exciton linewidth measurements (in Fig. 3f and Figs. S6 and S7 in Supplemental Material [22]) were obtained using two monochromators in cascade. The laser linewidth with two monochromators is ~0.8 meV. A 610 nm bandpass filter is kept in the detection side to monitor 1$s$ exciton emission. The average power used for the measurements in this work is $< 5\ \mu$W. Linear polarization resolved measurements are carried out by using a half-wave plate (analyzer) in the emission (detection) side. DOLP $\left[= (I_{H/H} - I_{\mathrm{H/V}})/\left(I_{H/H} + I_{\mathrm{H/V}}\right)\right]$ is obtained by keeping the incoming polarization in parallel $\left(I_{H/H}\right)$ and perpendicular $\left(I_{\mathrm{H/V}}\right)$ direction with respect to the analyzer. For the circular polarization resolved measurements, an additional quarter-wave plate is inserted right before the objective lens. DOCP $\left[= (I_{\sigma+/\sigma+} - I_{\sigma+/\sigma-})/\left(I_{\sigma+/\sigma+} + I_{\sigma+/\sigma-}\right)\right]$ is obtained by changing the incoming linear polarization direction from $+45^0$ to $-45^0$ with the fast-axis of the quarter-wave plate. For temperature dependent measurements, the excitation wavelength is tuned to ensure single-step resonance with the $A_1'$ phonon mode at each temperature.

**Acknowledgement**

This work was supported in part by a Core Research Grant from the Science and Engineering Research Board (SERB) under Department of Science and Technology (DST), a grant from SERB under TETRA, grants from Indian Space Research Organization (ISRO), a grant from MHRD under STARS, a grant from QuRP, IISc, and a grant from MHRD, MeitY and DST Nano Mission through NNetRA.

**Notes**

The authors declare no competing financial or non-financial interest.

**Data Availability**

Data available on reasonable request from the corresponding author.

**Figure 1: Exciton relaxation mechanisms:** Photons of energy $hc/\lambda_{ex}$ excites higher energy excitons inside the light cone of an $ns$ exciton band. These excitons can relax to the lowest energy $1s$ band in the following three ways: (a) M1 – through scattering with multiple phonons, (b) M2 – radiative transition through photon emission between the dipole-coupled Rydberg states, and (c) M3 – in a single-step by scattering with a resonating phonon mode of energy $\hbar\omega_{ph}(=\Delta_{ns\rightarrow 1s})$ . Finally, the $1s$ excitons inside the light cone radiatively recombine and emit a light of energy $hc/\lambda_{em}$.

**Figure 2: Excitation spectroscopy on GWG sample:** (a) Schematic representation of the GWG stack. (b) PL spectrum at 5 K taken with $532$ nm excitation on the stack. The PL shows the $1s$ and $2s$ exciton peak from monolayer $WS_2$, apart from the 2D ($*$) and G ($\cdot$) Raman mode from the encapsulating FLG flakes. Inset: Linewidth of the $1s$ exciton peak (9.5 $\pm$ 2.017 meV) obtained with off-resonant excitation. (c) Left Panel - Color plot of the excitation spectroscopy measurement showing the $1s$ exciton emission intensity (in log scale) as a function of the emission and excitation wavelength. Right Panel - The intensity (in linear scale) normalized at each excitation wavelength. The resonating phonon modes are labelled. The line cuts showing the emission spectrum at (d) $594.5$, (e) $583$ and (f) $572$ nm excitation are plotted to distinguish the emission from single-step relaxation M3 [green shaded curves in (d)] and two-step relaxation [green shaded curves in (e)] from multiple-step relaxation M1 [pink shaded curves in (d-f)].

**Figure 3: Reduction in inhomogeneous broadening due to dual resonance:** (a) Bottom panel: Spatial potential fluctuation due to inhomogeneity in the system (external + internal). Top panel: The resulting spatial fluctuation in the exciton energy (shaded curve). The intersections with the solid lines are the equi-energy spots at three different exciton energies (marked 1, 2 and 3). (b) Schematic description of the inhomogeneous PL broadening reduction obtained by single-step scattering with the $A_1'$ phonon mode due to dual-resonance. The bottom inset schematically shows the spatial profile of the equi-energy spots at three different $1s$ exciton emission energies [marked as 1, 2, and 3 in (a)]. (c) Left panel: PL broadening due to light emission from all the energy states (spots) in the distribution. Right panel: PL linewidth narrowing due to emission from specific $1s$ energy states resulting from narrow resonance in single-step relaxation. (d) Experimental result of single-step relaxation through $A_1'$ phonon mode (M3 process) in orange shaded curves, resulting in the sharp $1s$ exciton peak. The excitation wavelength is tuned such that the emission line scans through the inhomogeneously broadened PL emission peak (M1 process). The latter is obtained by off-resonant $532$ nm excitation (raw emission data in grey trace and fitting in black dashed trace). (e) Spatial imaging from $1s$ emission highlighting the change in the spatial distribution of the equi-energy spots at different exciton emission energies [marked by arrows in (e)]. The corresponding excitation wavelengths are indicated in the insets. (f) A linewidth of $\sim 0.19 \pm 0.07$ meV is obtained on deconvoluting the exciton spectrum (in symbols) with the excitation laser having nonzero broadening (in yellow shading). The fitted curve after deconvolution (in solid trace) is also shown

**Figure 4: Polarization control during single-step relaxation:** (a) During relaxation mechanism M1 and inside the light cone, the polarization degrades due to scattering coupled with pseudospin (dark purple arrow) precession around the exchange induced magnetic field (light purple arrow). (b) DOLP (in red) of the $1s$ exciton peak that forms after relaxation mechanism M1 (open red circles, obtained from pink background in Fig. 2d-f) and M3 (solid

red circles, obtained from green shaded peaks in Fig. 2d-f) is shown. The DOCP of the excitons formed after M3 relaxation process is also shown (solid blue circles). (c) Polarization resolved Raman spectrum of $WS_2$ in the GWG stack ($532$ nm excitation). The $A_1'$ phonon peak is fully polarized, indicating that the polarization of the incoming light is retained on scattering with the $A_1'$ phonons. (d-e) The result of single-step scattering with the $A_1'$ phonons leads to the observation of high (d) DOLP and (e) DOCP in the GWG stack. Inset: A schematic view of the mechanism. (f-g) The degradation in (f) DOLP and (g) DOCP due to scattering with the $2A_1'$ phonon mode. Inset: A schematic view of the mechanism. (h) Single-step scattering with a chiral phonon mode ($b$) leads to polarization reversal of the incoming light in the FLG-$MoS_2$-FLG stack (highlighted by grey shade). Inset: A schematic view of the mechanism.

**Figure 5: Temperature independent ultra-narrow, highly polarized exciton peaks:** (a) The exciton population distribution broadens with $k_B T$ during relaxation mechanism M1. This causes linewidth broadening due to enhanced (phonon assisted) contribution from the higher $|\mathbf{Q}|$ states outside the light cone. The higher $|\mathbf{Q}|$ states are also long-lived and experience larger exchange, leading to polarization degradation. (b) The population distribution is mainly limited within the light cone through relaxation mechanism M3. (c) Representative linear polarization resolved emission spectra at $100$, $150$, and $200$ K, obtained after single-step relaxation with the $A_1'$ phonon. The DOLP and deconvoluted linewidth values are shown in the insets. (d) The extracted $1s$ exciton PL linewidth (in black symbols) and the DOLP (in red symbols) as a function of temperature.

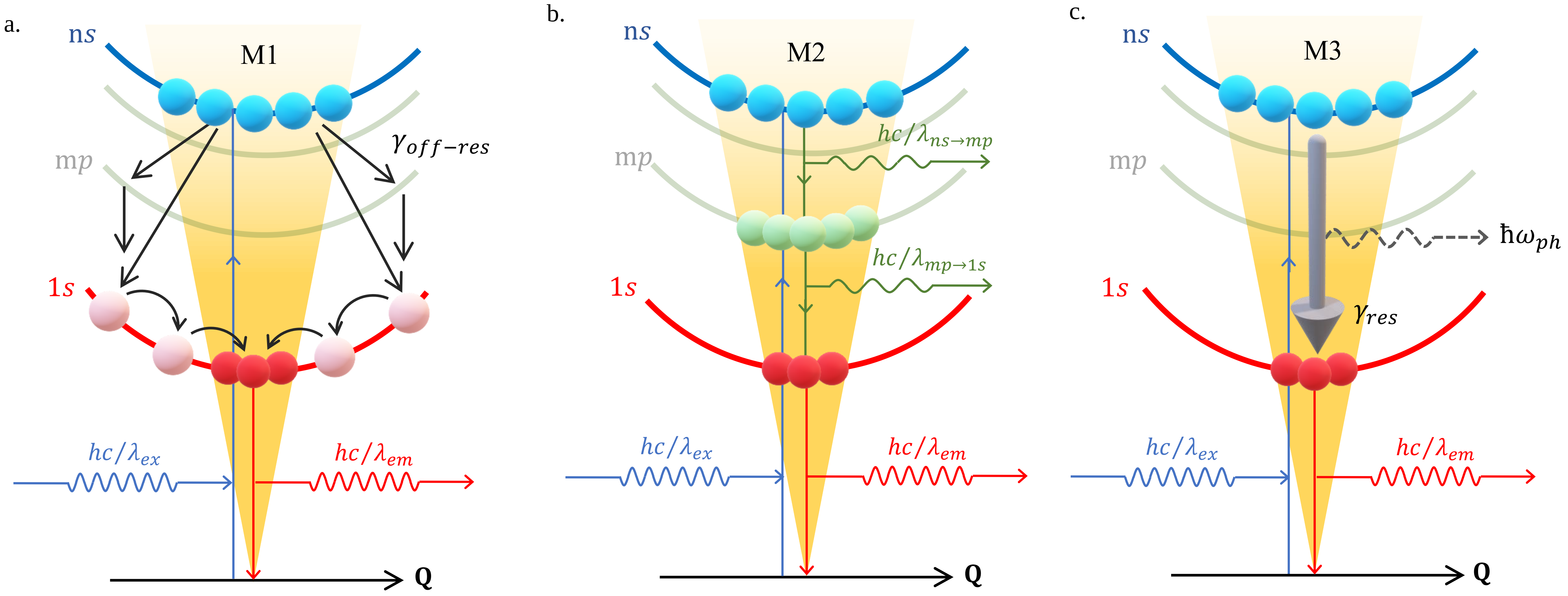

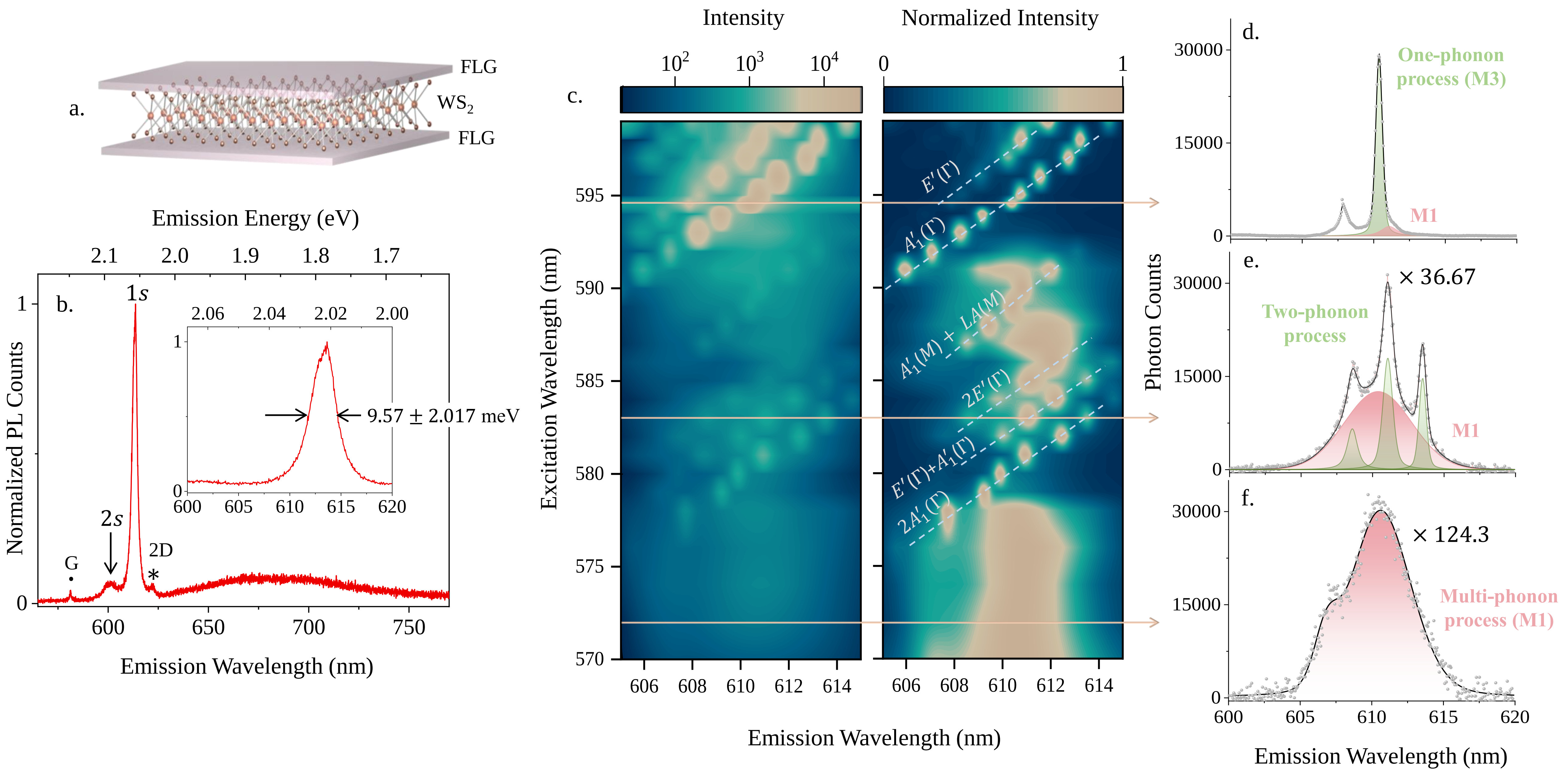

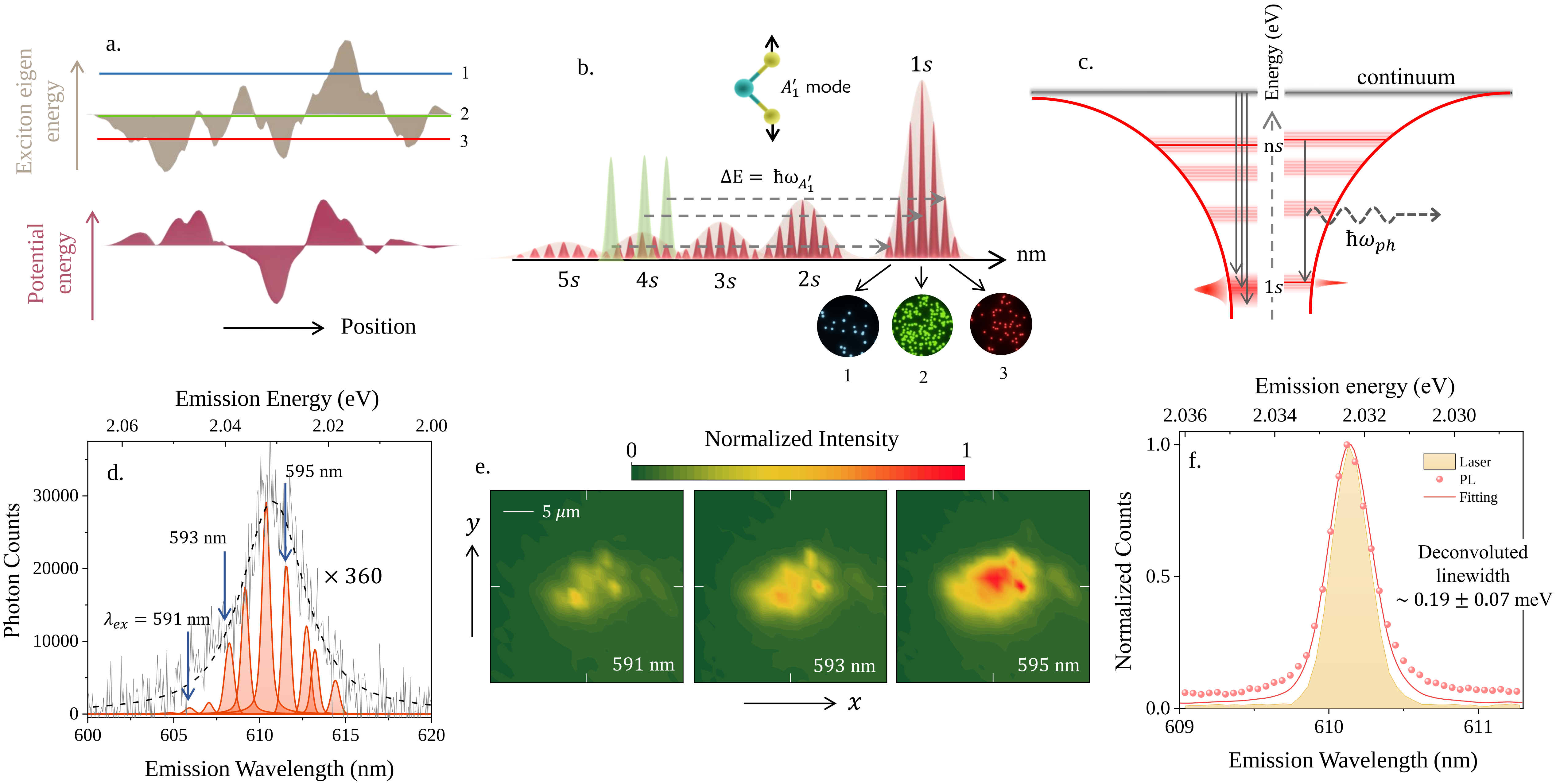

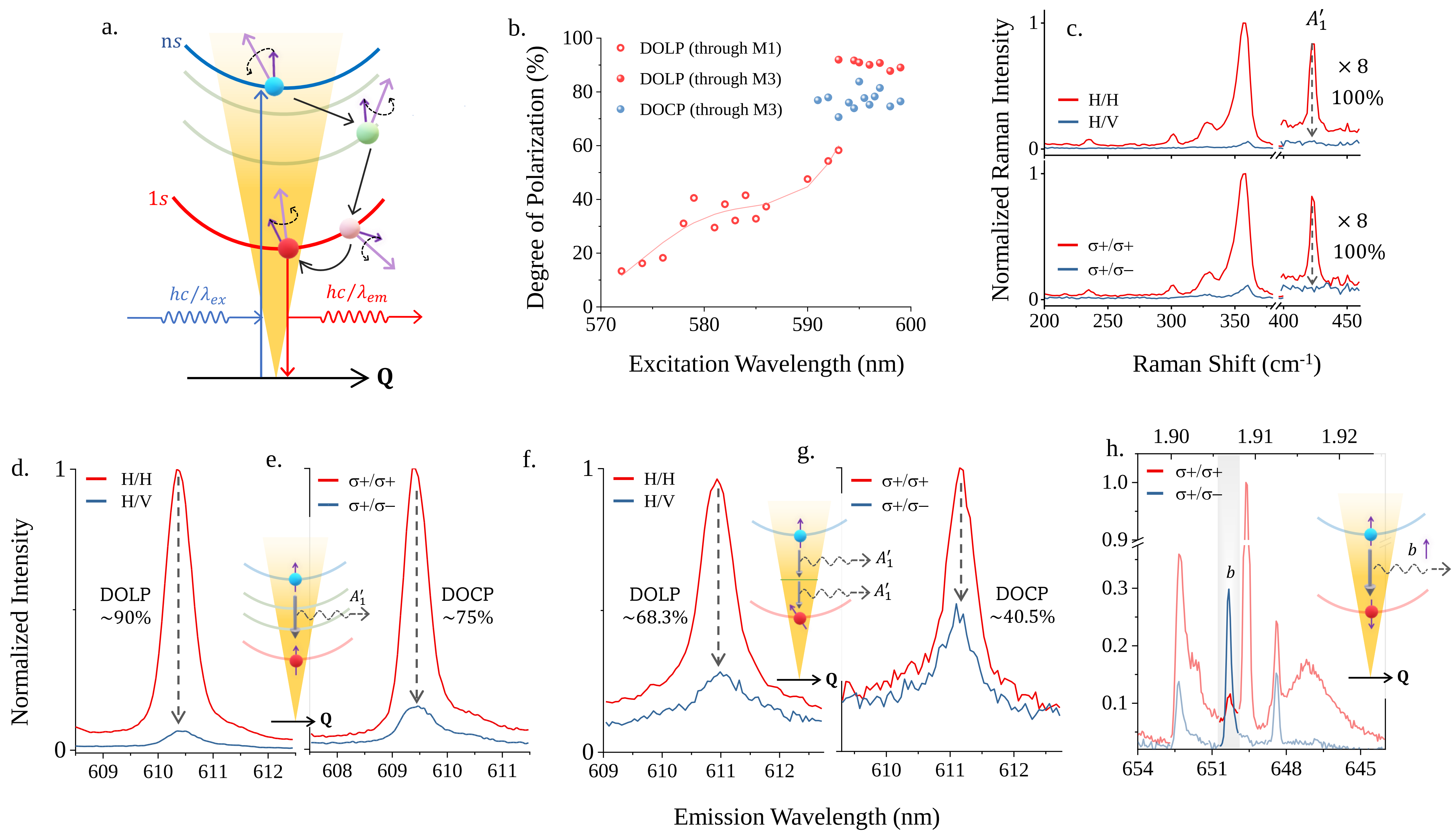

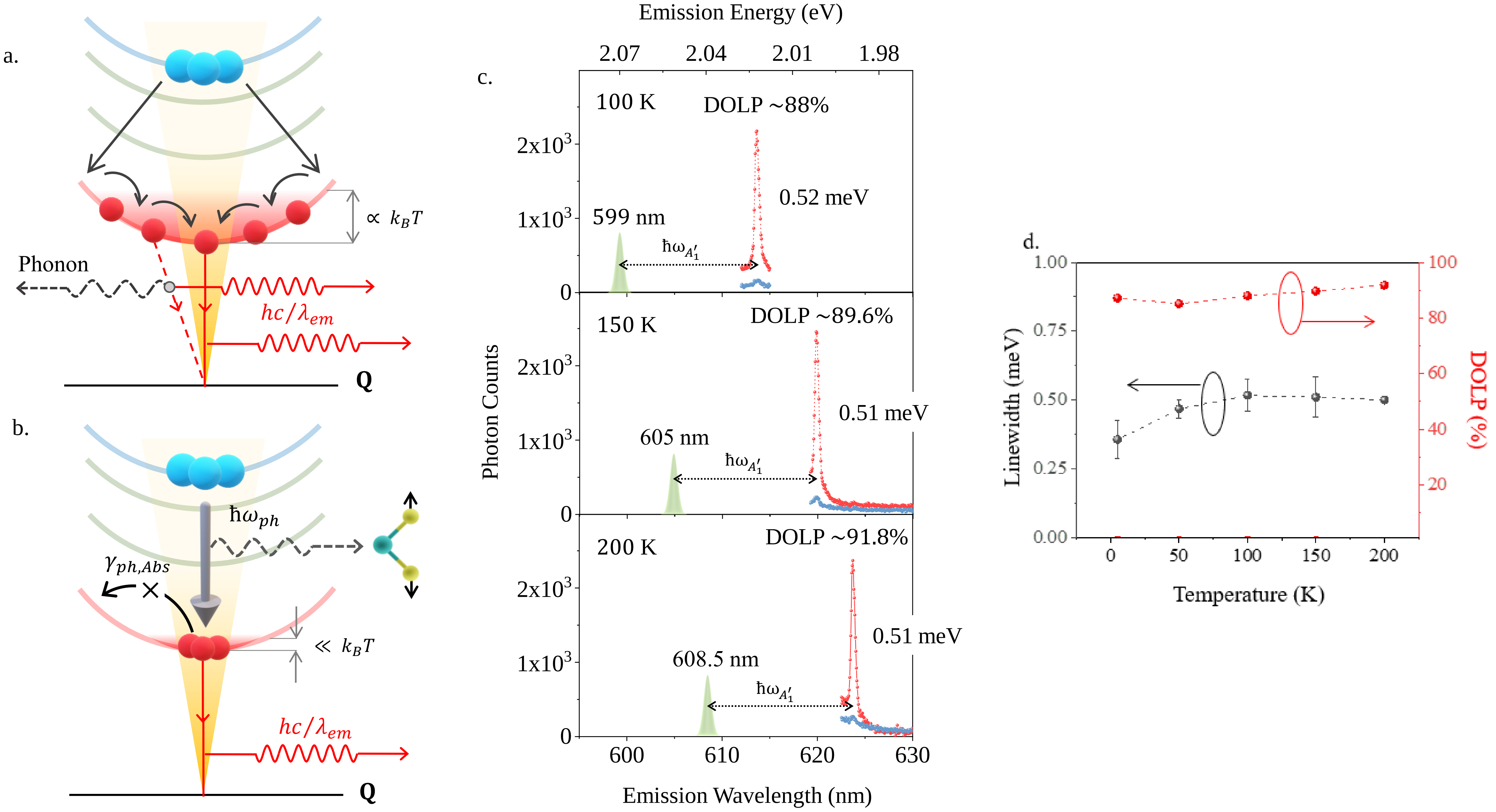

# Supplemental Material:

# Polarized and narrow excitonic emission from graphene-capped monolayer $WS_2$ through resonant phonon relaxation

Garima Gupta and Kausik Majumdar*

*Department of Electrical Communication Engineering, Indian Institute of Science, Bangalore 560012, India*

**Corresponding author, E-mail: kausikm@iisc.ac.in*

## Note S1. Bethe-Salpeter equation - Calculation details

Firstly, the dispersion of the conduction band (CB) and the valence band (VB) in the $\mathbf{K}, \mathbf{K}'$ valley of monolayer TMDs is calculated using the $2 \times 2$ **k.p** Hamiltonian in the basis states of $|d_{z^2}\rangle$ and $\frac{1}{\sqrt{2}}\left(|d_{x^2-y^2}\rangle + i\tau|d_{xy}\rangle\right)$, ($\tau$ (= $\pm1$) is the valley index) as given below [1]:

$$\begin{bmatrix} E_g & at(\tau k_x - ik_y) \\ at(\tau k_x + ik_y) & 0 \end{bmatrix}$$

$k_x, k_y$ represents the wave vectors in the reciprocal space. $a, t$ and $E_g$ is the lattice constant, hopping amplitude and the quasi-particle band gap for monolayer $WS_2$, respectively.

On solving the above matrix, we obtain the eigen energy (eigen wavefunction) of the CB and the VB, that we denote by $E_{c,\mathbf{k}}(|c,\mathbf{k}\rangle)$, $E_{v,\mathbf{k}}(|v,\mathbf{k}\rangle)$, respectively.

Secondly, to obtain the exciton eigen energy positions and wavefunctions, we solve the Bethe-Salpeter (BS) equation [2], which provides the exciton energy and wavefunction as a function of its centre-of-mass momentum [$\mathbf{Q} = (\mathbf{k_e} + \mathbf{k_h})$].

$$\langle vc\mathbf{kQ}|H|vc\mathbf{k}'\mathbf{Q}\rangle = \delta_{\mathbf{kk}'}\left(E_{c,\mathbf{k+Q}} - E_{v,\mathbf{k}}\right) - (D - X)(\mathbf{k}, \mathbf{k}', \mathbf{Q})$$

$|vc\mathbf{kQ}\rangle$ represents an electron ($\mathbf{k_e} = \mathbf{k}$+$\mathbf{Q}$) - hole ($\mathbf{k_h} = \mathbf{k}$) pair that forms as basis states of an exciton at $\mathbf{Q}$. $D$ is the direct coulombic interaction between the electron-hole pair states given by:

$$D = \frac{1}{A} V_{\mathbf{k}-\mathbf{k}'} \langle c, \mathbf{k} + \mathbf{Q}|c, \mathbf{k}' + \mathbf{Q}\rangle \langle v, \mathbf{k}'|v, \mathbf{k}\rangle$$

We neglect the exchange interaction $X$, as it has negligible contribution to the excitonic states close to $\mathbf{Q}$ = 0. $A$ is the area of the two-dimensional crystal. The interaction potential $V_{\mathbf{k}-\mathbf{k}'}$ is given by [3,4]:

$$V_q = \frac{2\pi {q_0}^2 e^{-q\xi}}{q} \frac{1}{\epsilon(q)}$$

Here $q_0$ is the elementary charge value, $q$ = $|\mathbf{k} - \mathbf{k}'|$, and $\xi$ is a fitting parameter denoting the spatial spread in the out-of-plane direction of an electron and hole. The dielectric function $\epsilon(q)$ is given by

$$\epsilon(q) = \frac{(1 - p_b p_t e^{-2\eta q D})\kappa}{(1 - p_t e^{-\eta q D})(1 - p_b e^{-\eta q D})} + r_0 q e^{-q\xi}$$

$p_b = p_t = P = (\epsilon_{ENV} - \kappa)/(\epsilon_{ENV} + \kappa)$ for identical dielectric environment on top and bottom. $D$ is the monolayer sheet thickness, $\eta = \sqrt{\epsilon_{\parallel}/\epsilon_{\perp}}$ and $\kappa = \sqrt{\epsilon_{\parallel}\epsilon_{\perp}}$. $\epsilon_{\parallel}(\epsilon_{\perp})$ and $\epsilon_{ENV}$ is the effective in-plane (out-of-plane) and environmental dielectric constant of the monolayer TMD. The value of $\epsilon_{\parallel}$ $(\epsilon_{\perp})$ used in our calculations is 5.01 (6.07).

To obtain the exciton eigen energy positions in a monolayer $WS_2$ flake encapsulated in graphene, we use the following fitting parameters to match the $\Delta_{2s-1s}$ with its experimentally obtained value of 31.5 meV:

| | |
|---|---|
| $\epsilon_{ENV}$ | 48 |
| $r_0$(nm) | 4 |
| $\xi$(nm) | 6 |

The excitonic eigen energy positions for the first few bright states are given in the table below:

| Eigen state | Eigen position (eV) |
|---|---|
| 1s | 2.03 |
| 2s | 2.062 |
| 3s | 2.075 |
| 4s | 2.082 |
| 5s | 2.087 |
| 6s | 2.09 |

## Note S2. Distinguishing excitonic emission from the resonant Raman process

The photons detected at the dual resonance (both absorption and emission resonance) can be obtained through the two possible channels:

(a) Resonant Raman scattering: Here the excitation is resonant with the higher energy excitonic states, and the final state after Raman scattering is a vibrational state above the ground state (see Figure S2.1, left panel).

(b) Emission process – A cascade of photon absorption, relaxation of the generated ns exciton to 1s state by phonon emission, followed by light emission from the 1s exciton (see Figure S2.1, right panel): In this work, the main motivation behind laser excitation near an optical phonon resonance is the following – When the higher energy excitons are generated, they relax to the light cone of the lowermost 1s exciton band in a single step, through scattering with a particular resonating phonon mode (Figure 1c in the main text). We refer to this single-step relaxation mechanism as the M3 mechanism throughout the manuscript. This is done to avoid the conventional energy relaxation process, which takes place through scattering with multiple phonons Figure 1a in the main text) - referred to as the M1 mechanism throughout the manuscript.

The two processes (a and b) are schematically shown and compared in the Figure S2.1. Here we ignore the contribution from the relatively weaker processes that involves virtual states.

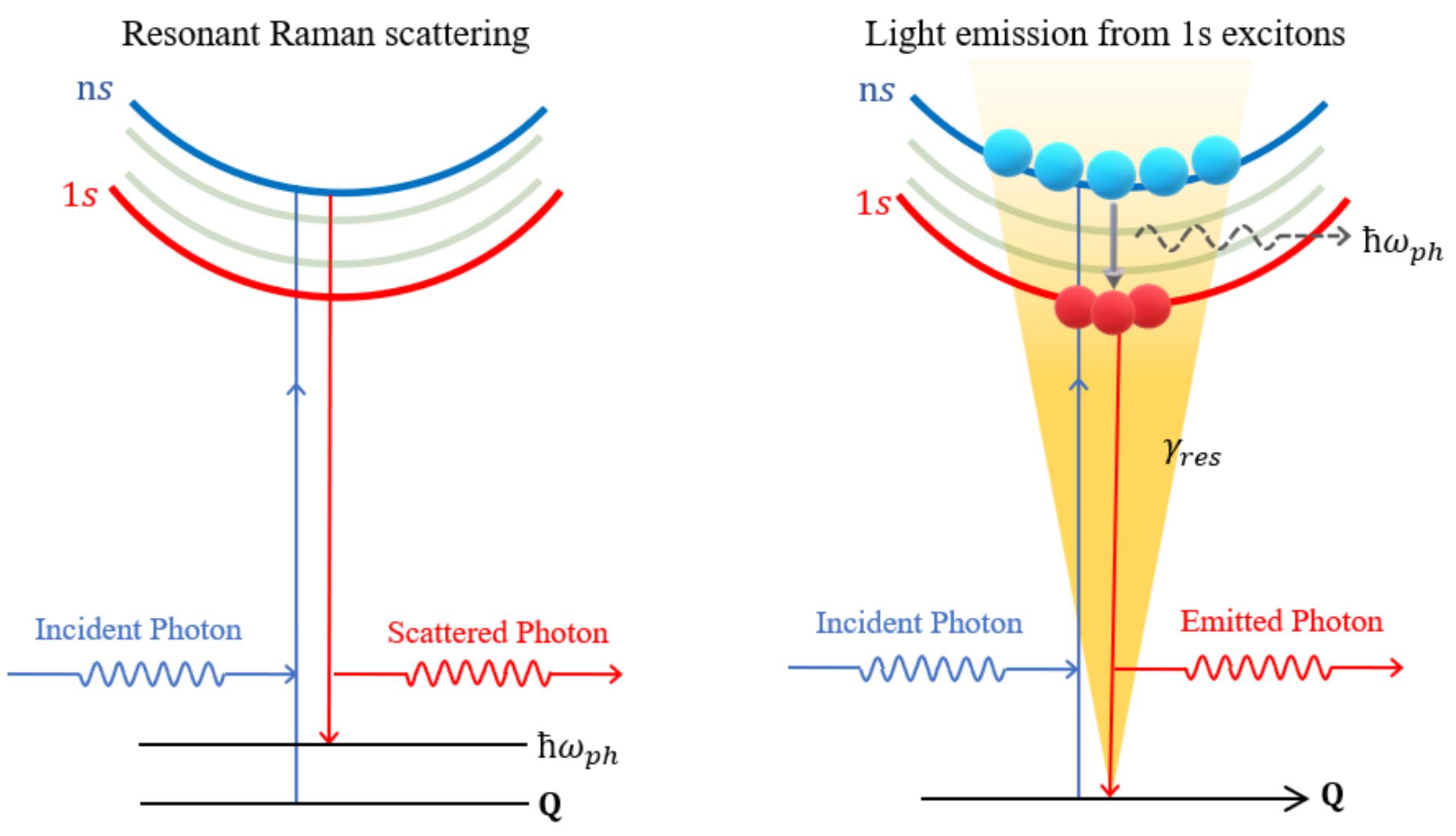

Figure S2.1: Schematic showing resonant Raman scattering (left panel) and exciton emission after single-step resonant phonon relaxation.

In general, both the Raman scattered photon, and the emitted photon from radiative recombination of the 1s exciton will contribute to the detected light (at 610 nm in our case). Note that, Resonant Raman scattering is an instantaneous process that maintains complete coherence with the excitation source, while the light detected due to the emission process results from a cascade of absorption, exciton scattering by phonon, and emission [5,6]. The emission of the 1s exciton has a characteristic lifetime.

While both processes can contribute to the outgoing photons, based on the experimental evidence, it is clear that the emission from 1s excitons (process b) is the dominant mechanism in our case as compared to the resonant Raman scattered light (process a). The arguments are as follows:

(1) **Difference in the polarization dependence of the Raman signal and the excitonic emission**

As stated above, in the Resonant Raman process, a complete coherence with the excitation source is maintained and the scattered light can be fully polarized (same as the excitation polarization) if the scattering phonon mode has a diagonal Raman tensor form, for example, the $A_1'$ phonon mode. This is shown in Figure S2.2 in the polarization resolved Raman spectrum of $WS_2$ showing a 100% degree of linear (top panel) and circular (bottom panel) polarization of the Raman scattered light with the $A_1'$ phonon.

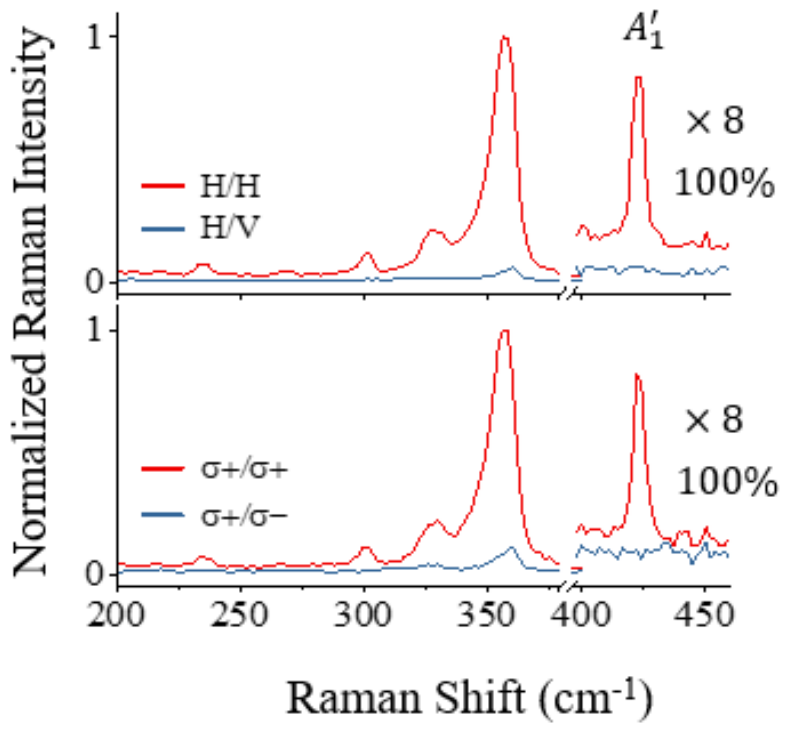

Figure S2.2: Linear (top panel) and circular (bottom panel) polarization resolved Raman spectrum of monolayer $WS_2$.

Had Resonant Raman been the dominant mechanism, we should expect the output polarization to be 100% linearly and circularly polarized in our case as well. However, we obtain a value well below 100% degree of linear polarization (~90%), and an even lower degree of circular polarization (~75%), when the excitation laser is tuned such that $A_1'$ phonon is the resonating phonon mode during the single-step M3 relaxation process (Figure S2.3).

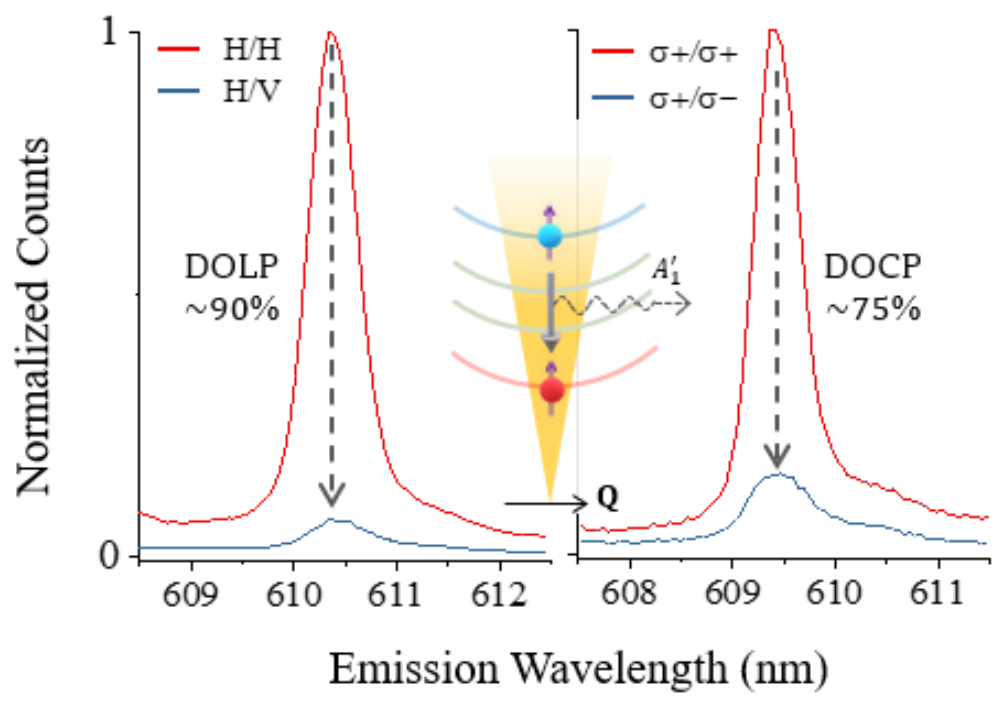

Figure S2.3: Linear (left panel) and circular (right panel) polarization resolved exciton emission spectrum on resonant relaxation through $A_1'$ phonon mode.

The observation of below 100% emission polarization as compared to the 100% polarization of the pure Raman scattered light can be explained by the role of MSS mechanism of pseudospin precession in the intermediate excitonic state, explained as follows – Firstly, since the polarization of the incoming light does not change by scattering with the $A_1'$ phonon mode due to the diagonal form of its Raman tensor, pseudospin randomization during relaxation through M3 process is evaded (Figure 4a in the main text). Secondly, during the finite stay of the excitons inside the light cone, the polarization of the relaxed excitons suffers due to pseudospin precession around the in-plane magnetic field manifesting due to large electron-hole coulomb interaction in monolayer TMDs. This leads to polarization degradation in the exciton band, as observed in our measurements.

This effect is even more pronounced in the case of resonance with the $2A_1'$ phonon mode. Again, as Raman scattering is not a cascaded (rather, an instantaneous and coherent) process, we expect 100% polarization of the Raman scattered light with the $2A_1'$ phonon. But we obtain a degraded DOLP (DOCP) of ~68.3 % (~40%) (see Figure S2.4). We believe that the enhanced degradation in this case as compared to the case of $A_1'$ phonon resonance is due to the additional exchange interaction suffered by the excitons in the intermediate state between two consecutive scatterings with the $A_1'$ phonon during relaxation to the 1s state (see inset in the Figure S2.4).

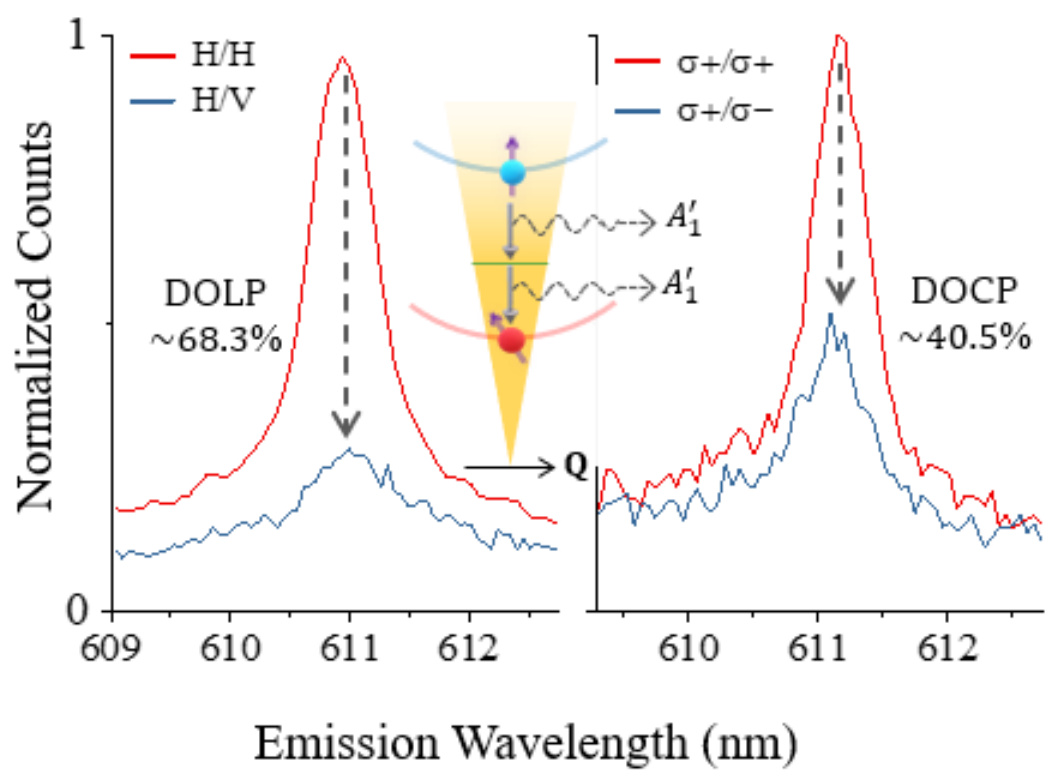

Figure S2.4: Linear (left panel) and circular (right panel) polarization resolved exciton emission spectrum on resonant relaxation through $2A_1'$ phonon mode.

Thus, the strong deviation from 100% polarization is a direct proof [6] that the emission process is more dominant than the resonant Raman process in our sample.

(2) **DOCP < DOLP – A characteristic feature of exciton pseudospin in 2D materials**

In 2D materials, due to the in-plane nature of the magnetic field manifesting due to electron-hole exchange interaction, it is expected that the degree of circular polarization is always lesser than the degree of linear polarization.

This is however not the case with the Raman scattered light (100% DOLP and DOCP of the Raman scattered light with $A_1'$ phonon, shown above).

As explained in the previous point, we observe DOCP < DOLP for both $A_1'$ and $2A_1'$ phonon resonance, again suggesting that emission playing a more dominant role than resonant Raman.

(3) **Distinguishing excitonic luminescence from Raman scattered light through change in the $WS_2$ layer number**

Another proof of strong excitonic influence is the intensity variation of the detected light on changing the excitation wavelength. We observe that the sharp peaks nicely trace the 1s PL envelope (obtained with off-resonant 532 nm excitation) on the stack of monolayer $WS_2$ encapsulated with FLG encapsulation (shown in Figure S2.5 and also shown in Fig. 3d in the main text). Such excellent envelope tracing of the 1s state clearly indicates that in the case of dual resonance, the 1s state plays an important role. This is another proof that the 1s emission process is the dominant mechanism.

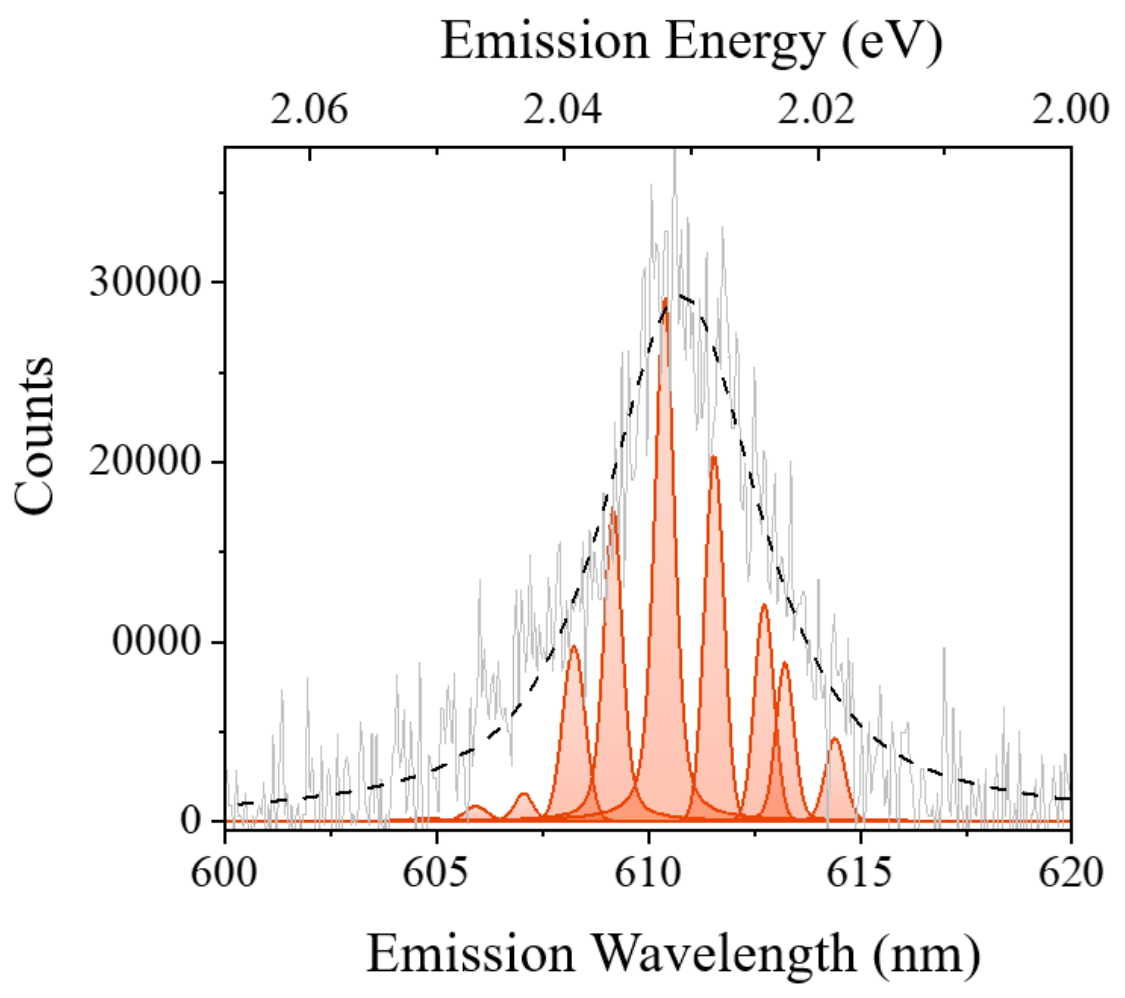

Figure S2.5: Spectra obtained from FLG encapsulated monolayer $WS_2$ using off-resonant (grey, multiplied by 360 in the plot) and resonant (red shading) excitations. The envelope is the PL spectrum taken with 532 nm excitation laser.

In contrast, when we apply the same technique to a similar stack of bilayer $WS_2$ flake encapsulated in few-layer-graphene (Figure S2.6), we do not observe such envelope tracing effect in the light output. Rather, in stark contrast to the monolayer case, the light intensity is found to be nearly independent of the energetic position of the excitation with respect to the inhomogeneously broadened excitonic emission shape (obtained from 532 nm off-resonant excitation):

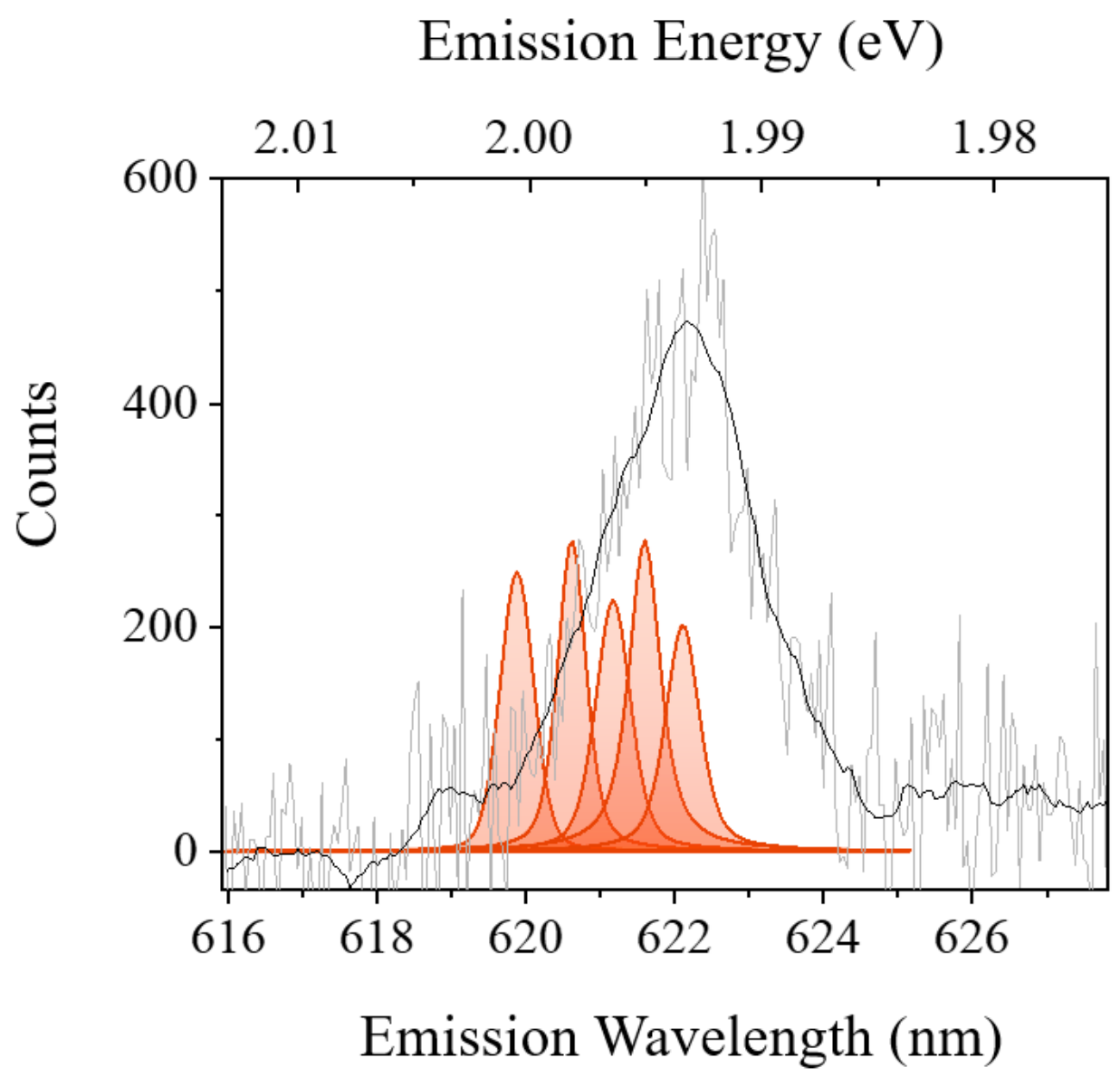

Figure S2.6: Spectra obtained from FLG encapsulated bilayer $WS_2$ using off-resonant (grey) and resonant (red shading) excitations. We do not observe any resonance feature like the monolayer case.

Such pattern is again observed in a four-layer (Figure S2.7) and five-layer $WS_2$ (Figure S2.8) flake encapsulated in graphene as well.

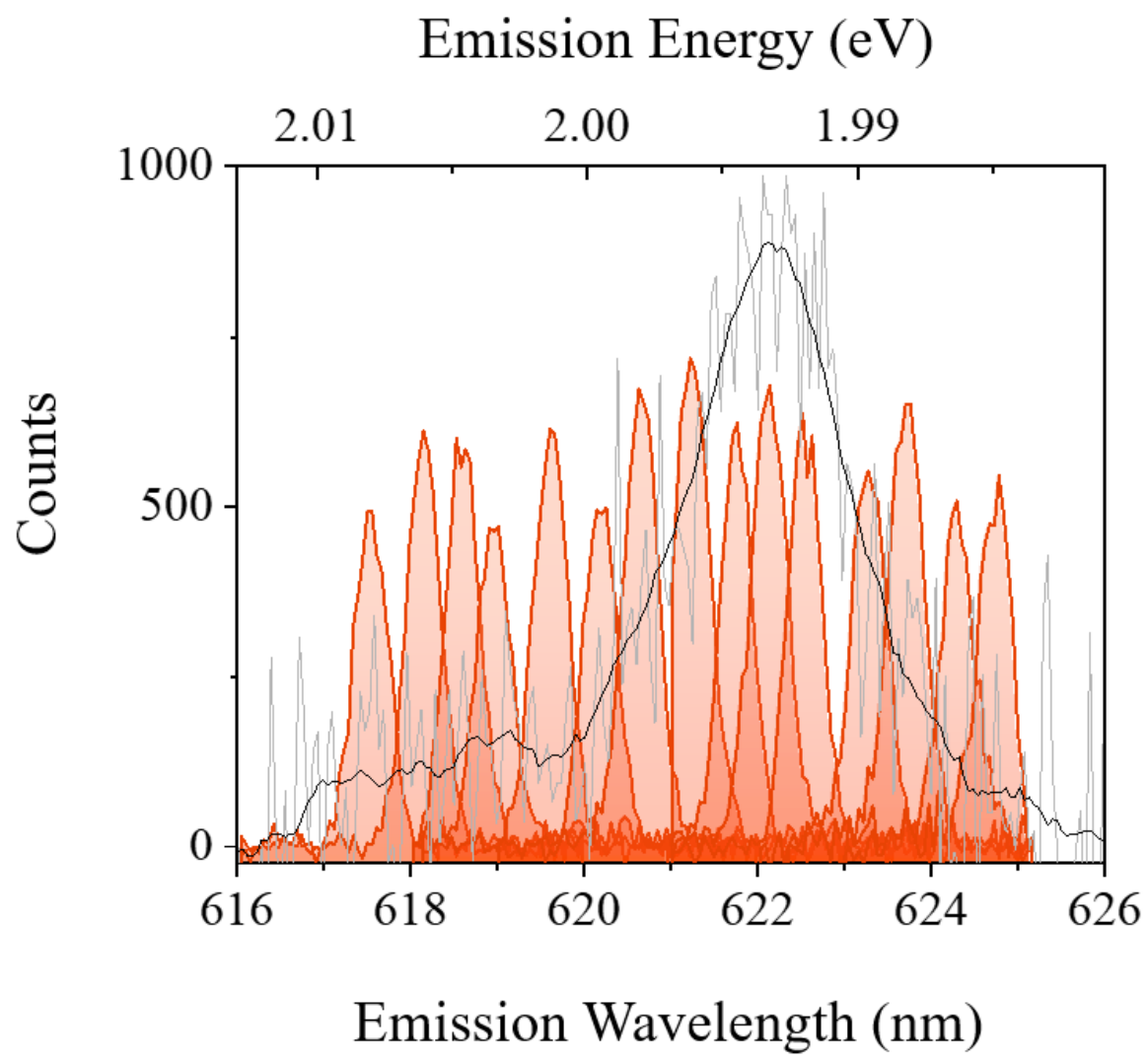

Figure S2.7: Spectra obtained from FLG encapsulated four-layer $WS_2$ using off-resonant (grey) and resonant (red shading) excitations. We do not observe any resonance feature like the monolayer case.

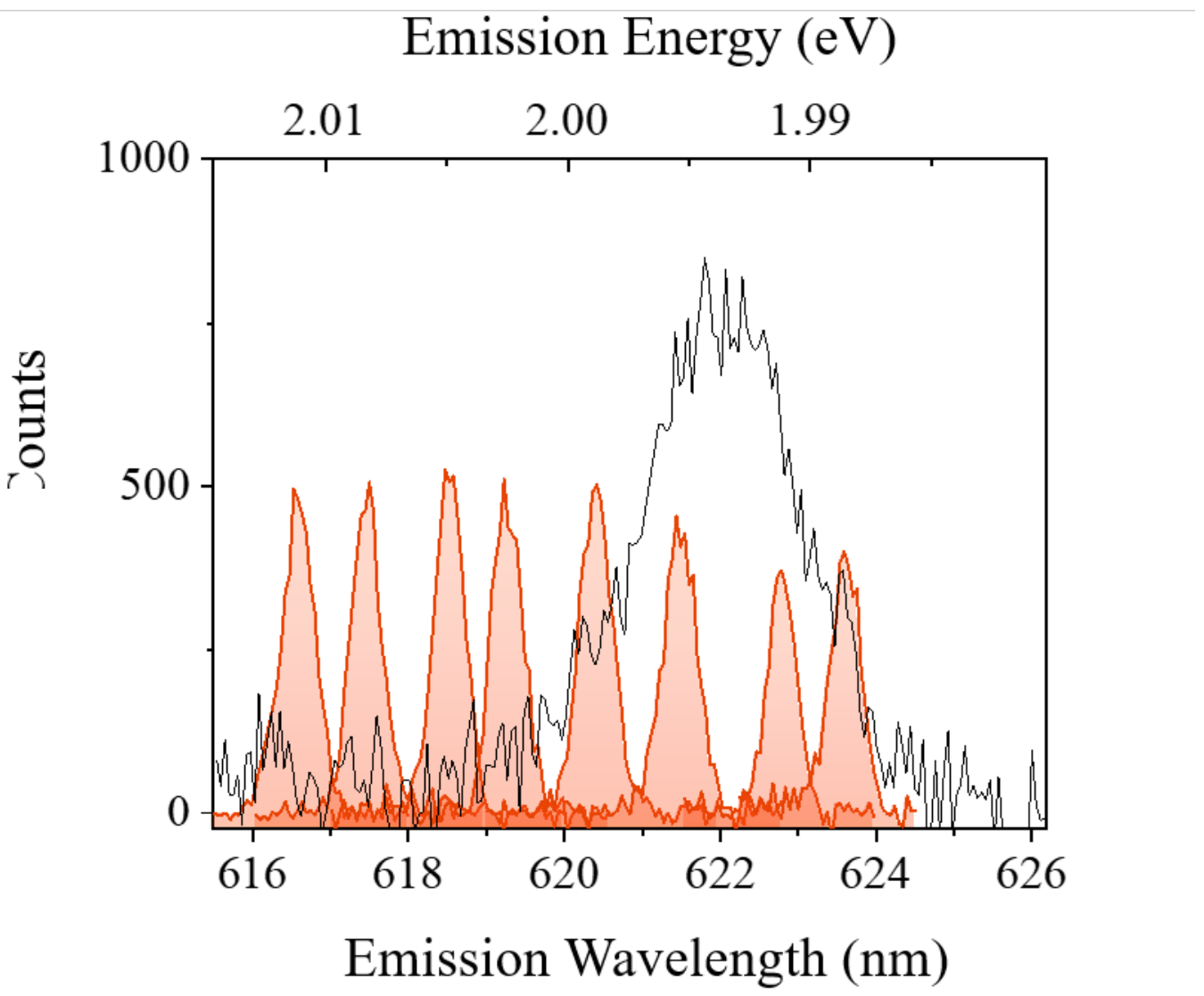

Figure S2.8: Spectra obtained from FLG encapsulated five-layer $WS_2$ using off-resonant (grey) and resonant (red shading) excitations. We do not observe any resonance feature like the monolayer case.

This difference between the monolayer and the few-layer $WS_2$ is explained using the internal quantum efficiency equation as given in the main text.

$$\eta_{int} = \underbrace{\frac{\gamma_{res}}{\gamma_{res} + \gamma_{off-res} + \gamma_{gr}}}_{\eta_{rel}} \times \underbrace{\frac{\gamma_r}{\gamma_r + \gamma_{nr} + \gamma_{gr}}}_{\eta_{em}}$$

Here $\eta_{rel}$ denotes the quantum efficiency of the relaxation process and $\eta_{em}$ denotes the quantum efficiency of the emission process once the $1s$ excitons are inside the light cone. It is known that as we move beyond a monolayer system, there are additional lower energy dark states, which increase linearly in number with the layer number ($L$) [7]. Therefore, fast scattering to these additional states significantly enhances the non-radiative decay rate ($\gamma_{nr}$) for $L \geq 2$, which results in degraded $\eta_{em}$. Therefore, despite the dual resonance, we do not observe strong excitonic emission through process (b), unlike in a monolayer case. The lack of any resonance feature suggests that the sharp peaks obtained for $L \geq 2$ during the wavelength scan, as shown above, do not depend on the 1s exciton state and are merely Raman scattered light, which do not carry any excitonic recombination information.

(4) **Abrupt reduction in peak intensity for $L > 1$**

From the above figures, we observe a significantly reduced peak intensity for $L > 1$ compared with monolayer samples. If monolayer peak intensity would have been dominated by resonant Raman process, we would expect similar intensity in the few-layer samples as well. This is another evidence that the peak intensity in the monolayer sample is dominated by excitonic emission under dual resonance, and not by resonant Raman scattering.

(5) **Distinction between Raman peak and excitonic emission peak linewidth with respect to the excitation laser**

The linewidth of the off-resonant $A_1'$ Raman peak in our GWG sample is very similar to the linewidth of the excitation laser (see left panel in Figure S2.9), which is expected for a Raman scattered signal. However, we obtain a significant difference between the laser linewidth and the excitonic emission linewidth in our measurement (see right panel in Figure S2.9). Please note that the noticeable difference between the laser linewidth ($\sim$ 0.8 meV) and the exciton peak linewidth $\sim$ 1.06 meV in the latter case is well above our system resolution of $\sim 40 - 60$ μeV.

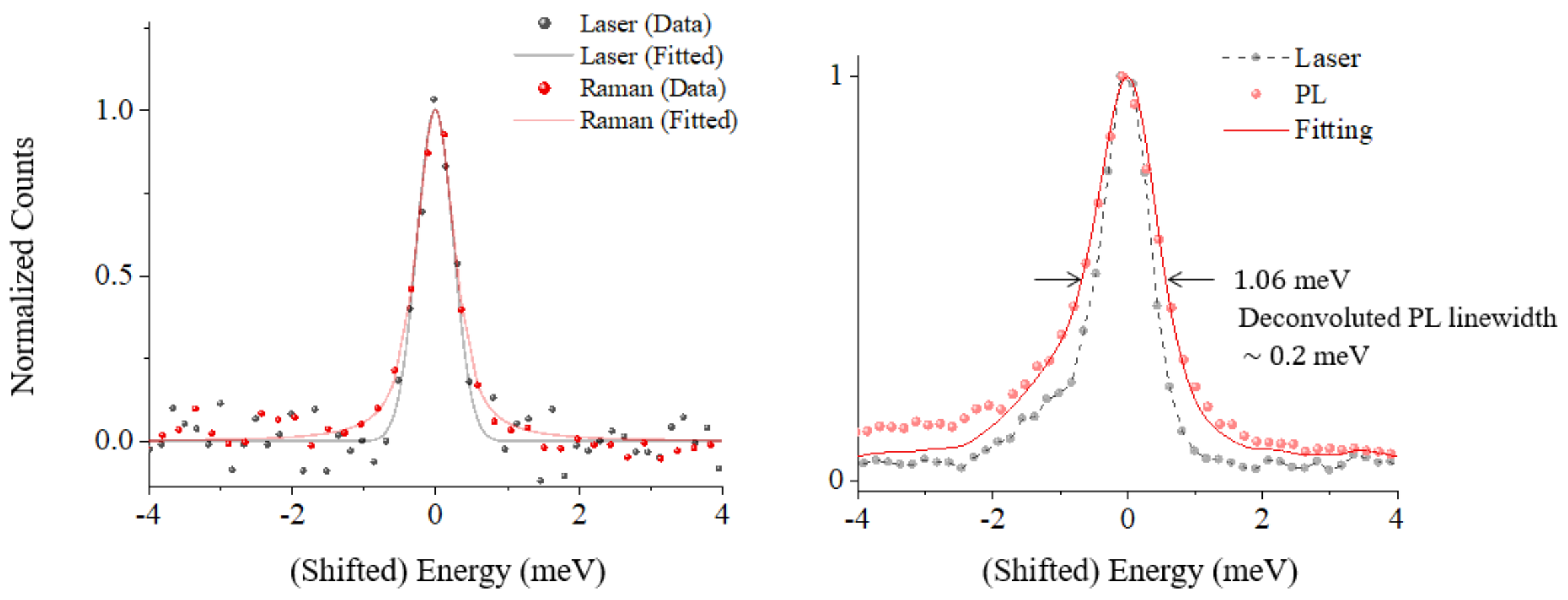

Figure S2.9: Left panel: Laser profile (data in black symbols, fitting in black solid trace) and the $A_1'$ Raman peak signal (data in red symbols, fitting in red solid trace) obtained with off-resonant excitation in GWG sample. Both almost completely overlap with each other due to similar linewidth. Right panel: Laser profile (data in black) and the exciton emission (data in red symbols, final fitted peak after laser convolution in red solid trace) through the M3 process on resonant excitation in our measurement in the GWG sample.

**Fig. S3:** Linear polarization resolved spectrum at each excitation wavelength (from main text color plot in Fig. 2c). The co- and the cross-polarized signal are shown in black and red, respectively. In each plot below, the vertical axis represents photon counts and the horizontal axis represents emission wavelength in nm.

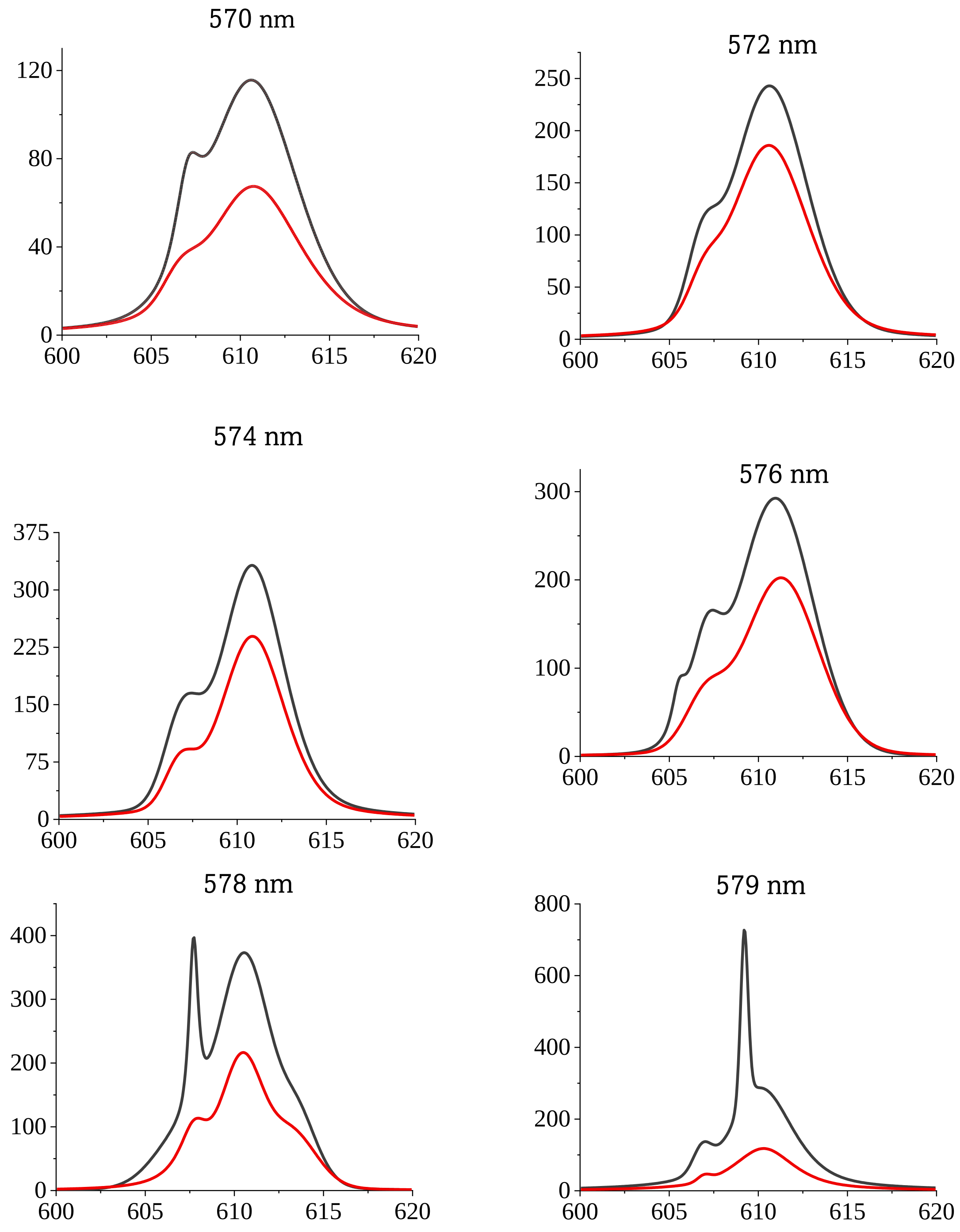

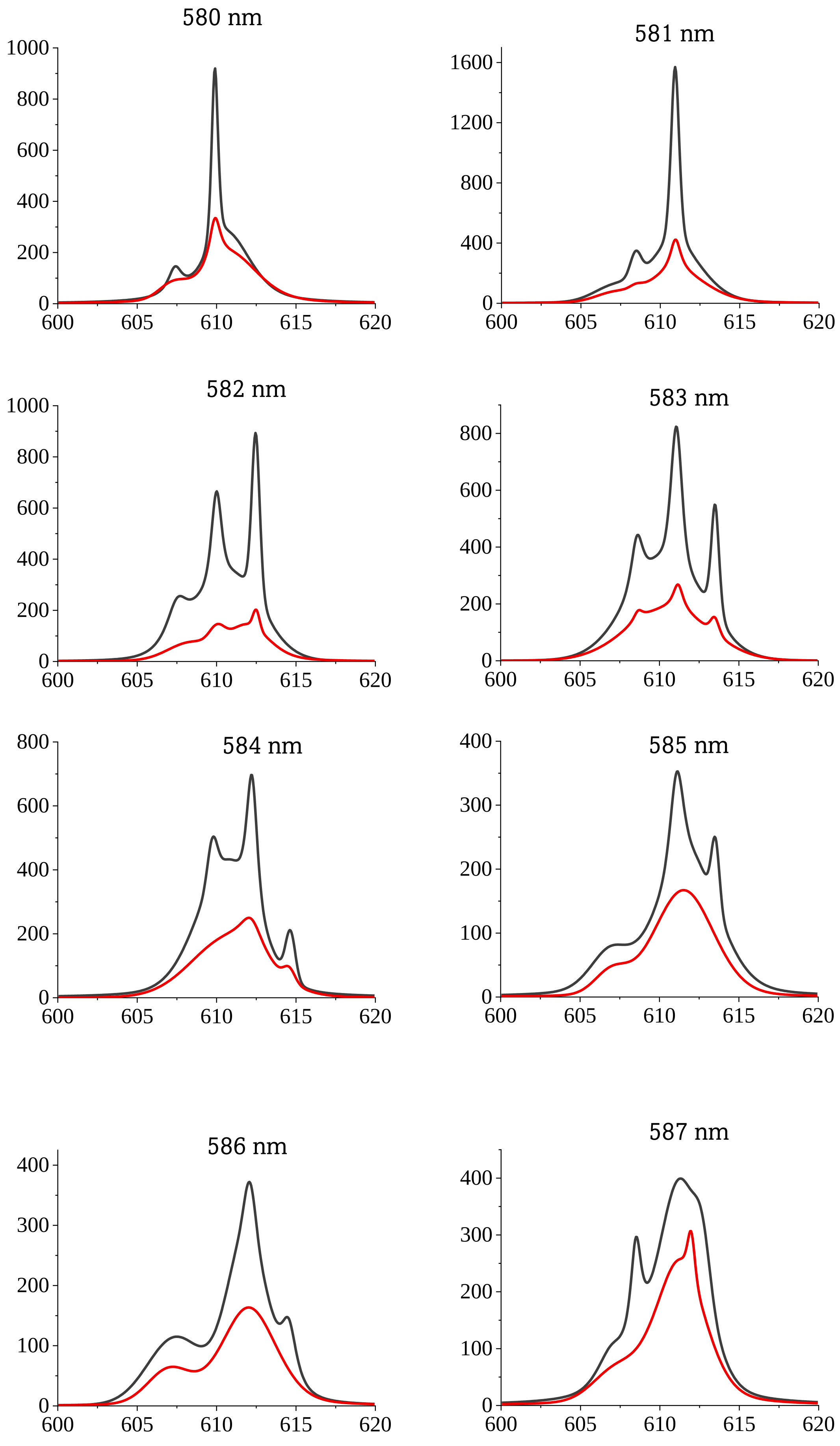

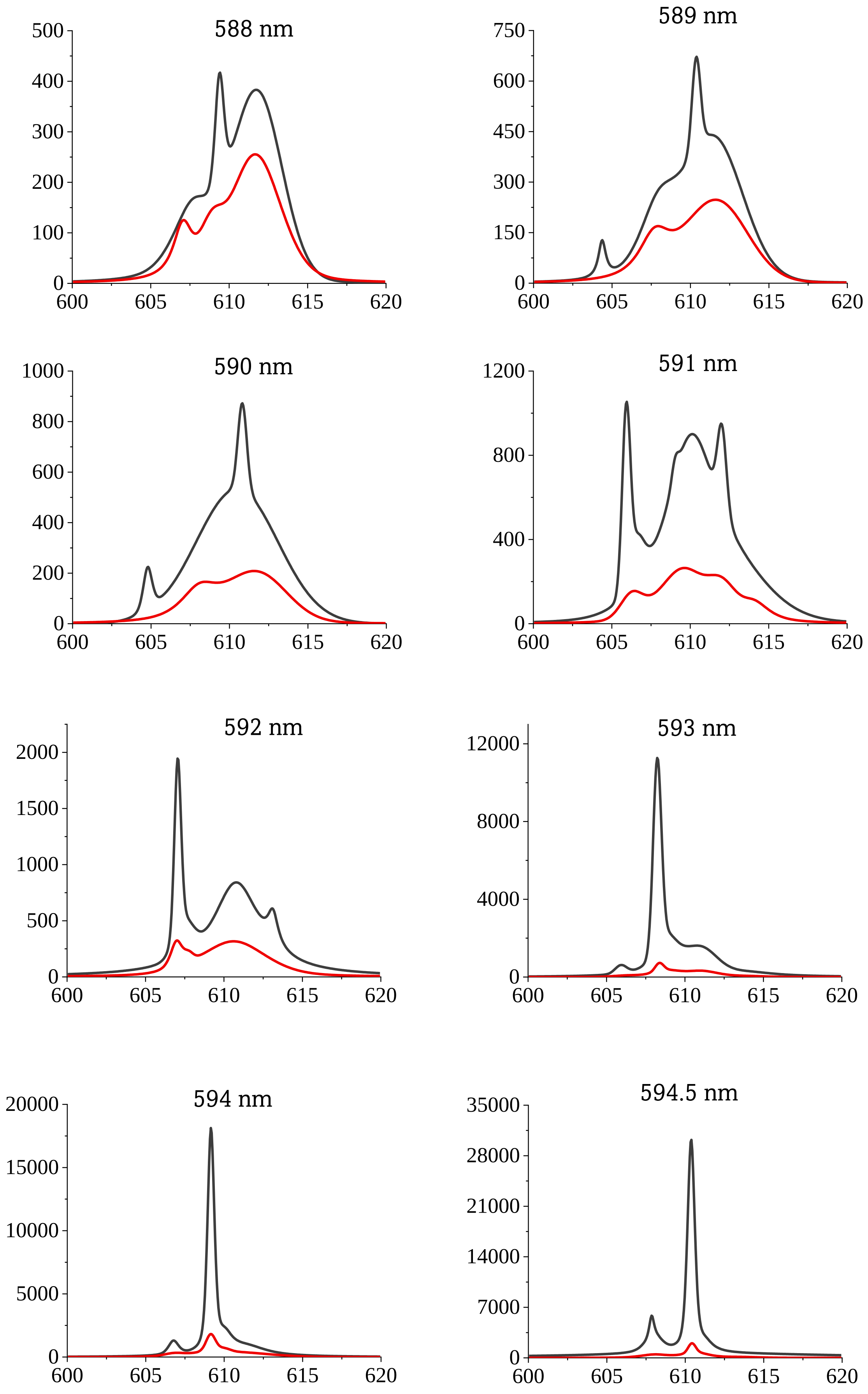

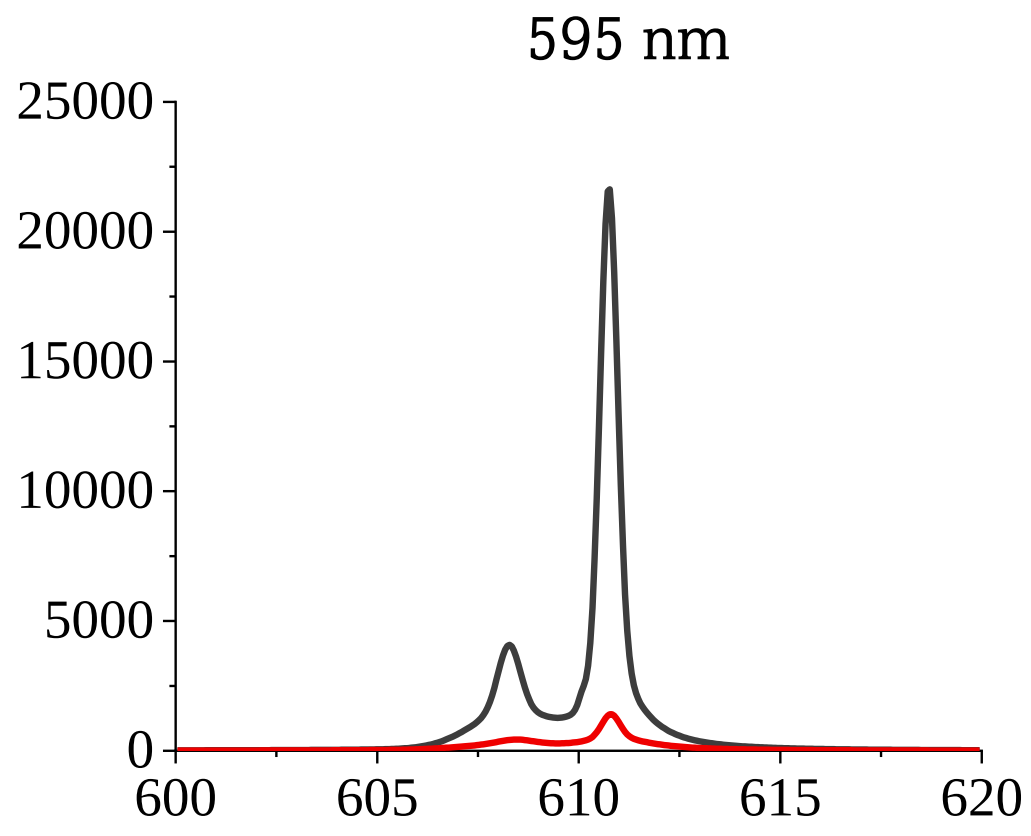

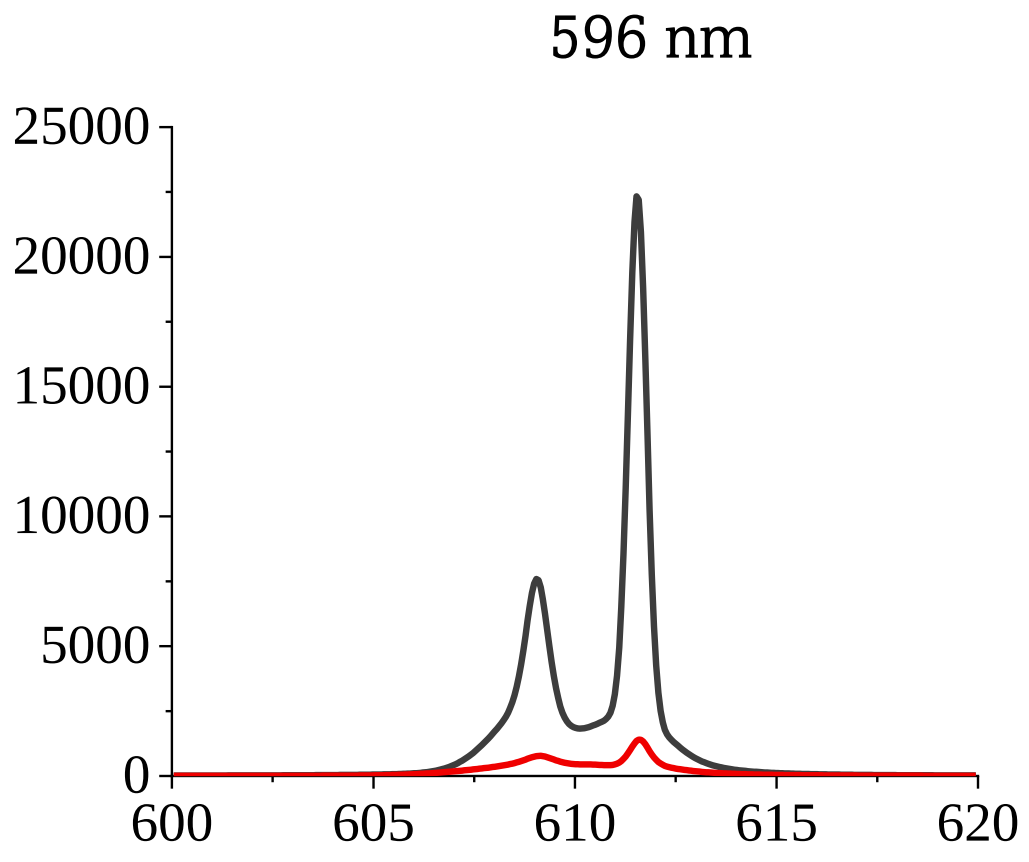

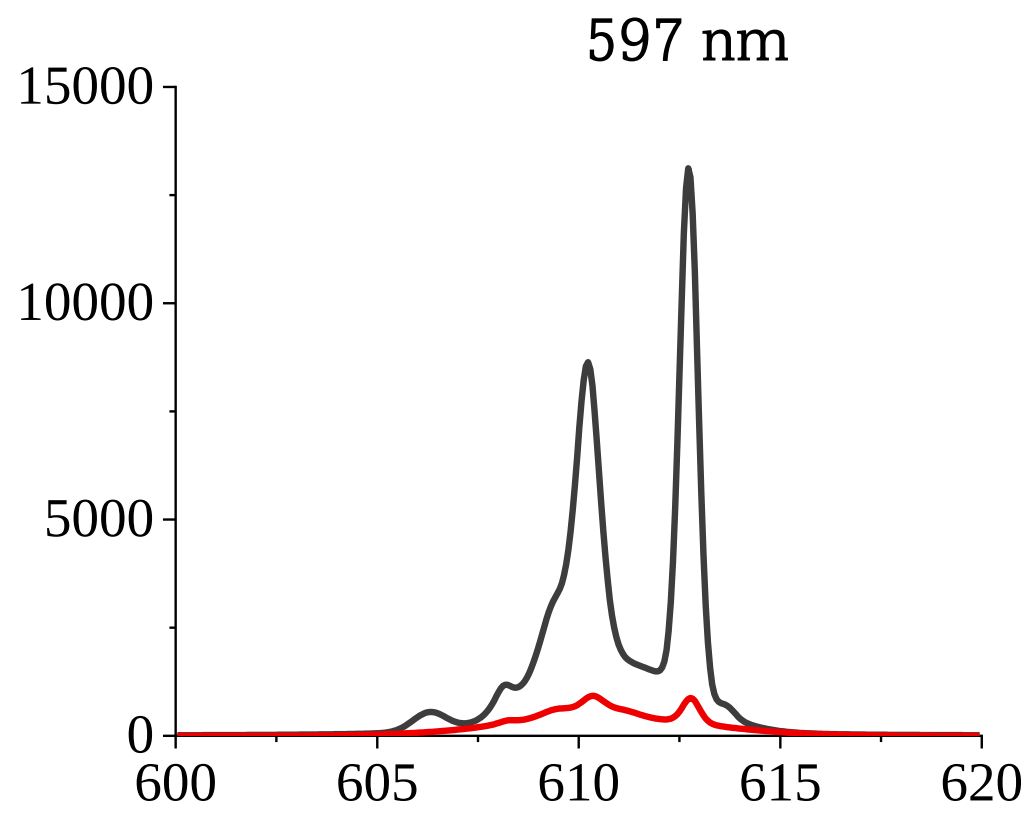

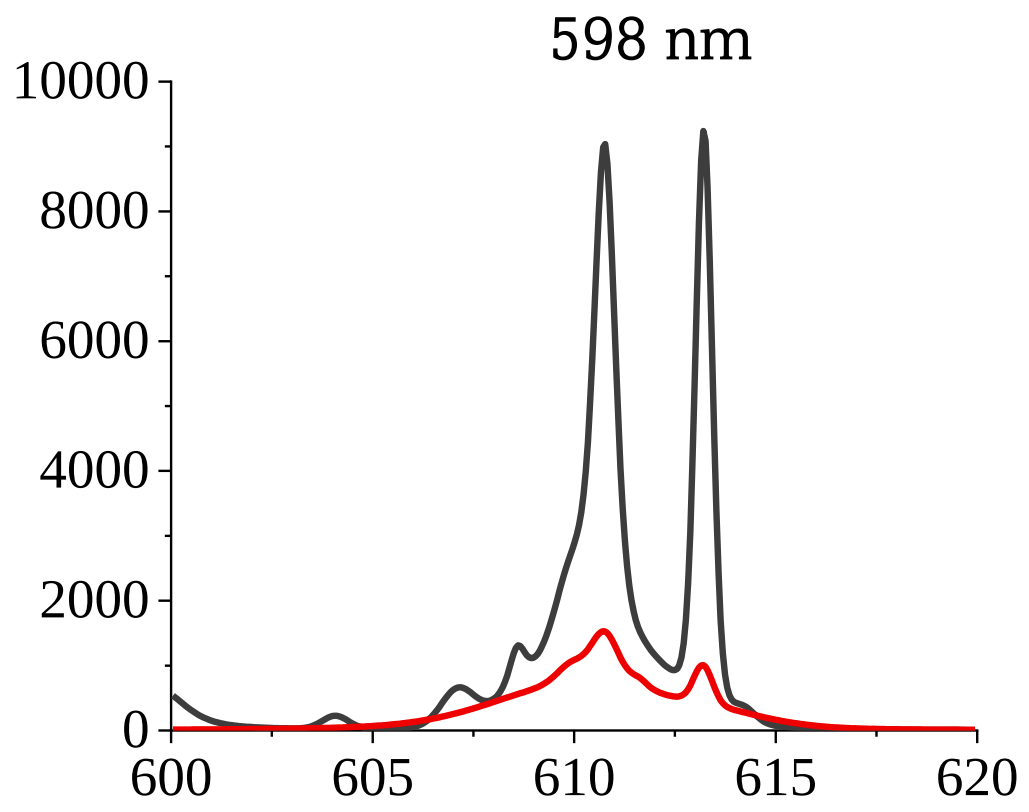

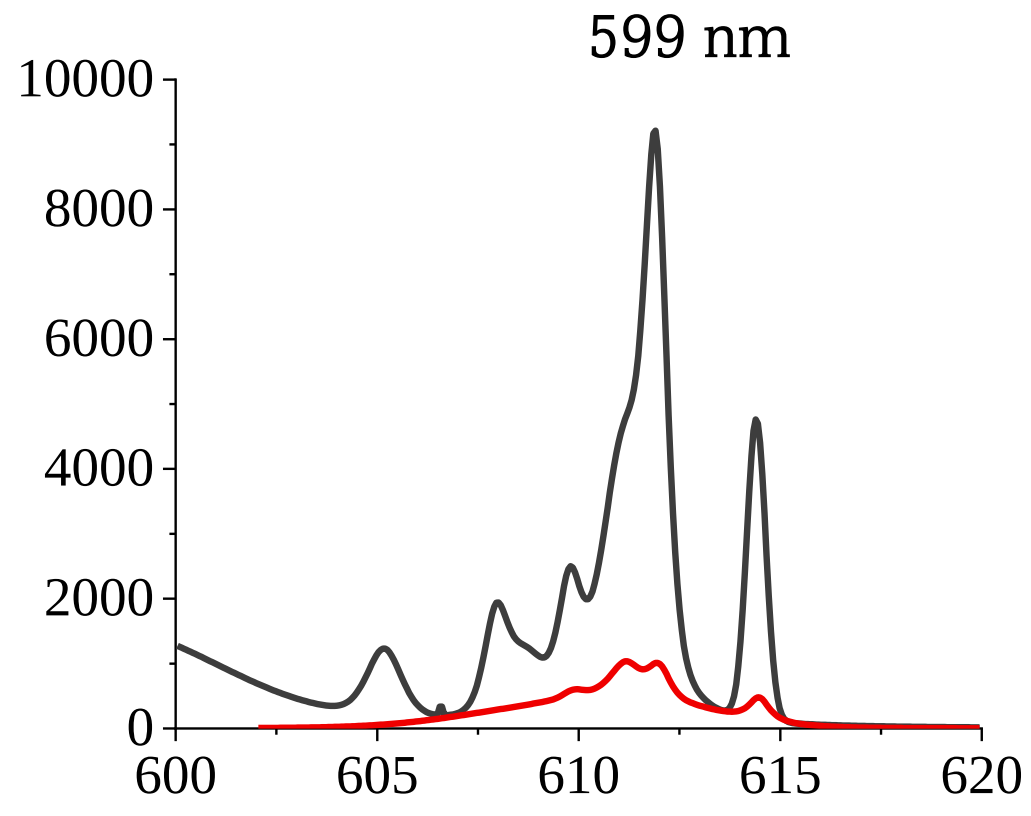

**Fig. S4:** Linewidth (after deconvolution with the excitation laser) of the narrow exciton peaks, in the main text Fig. 3d (in red shade) as a function of excitation wavelength.

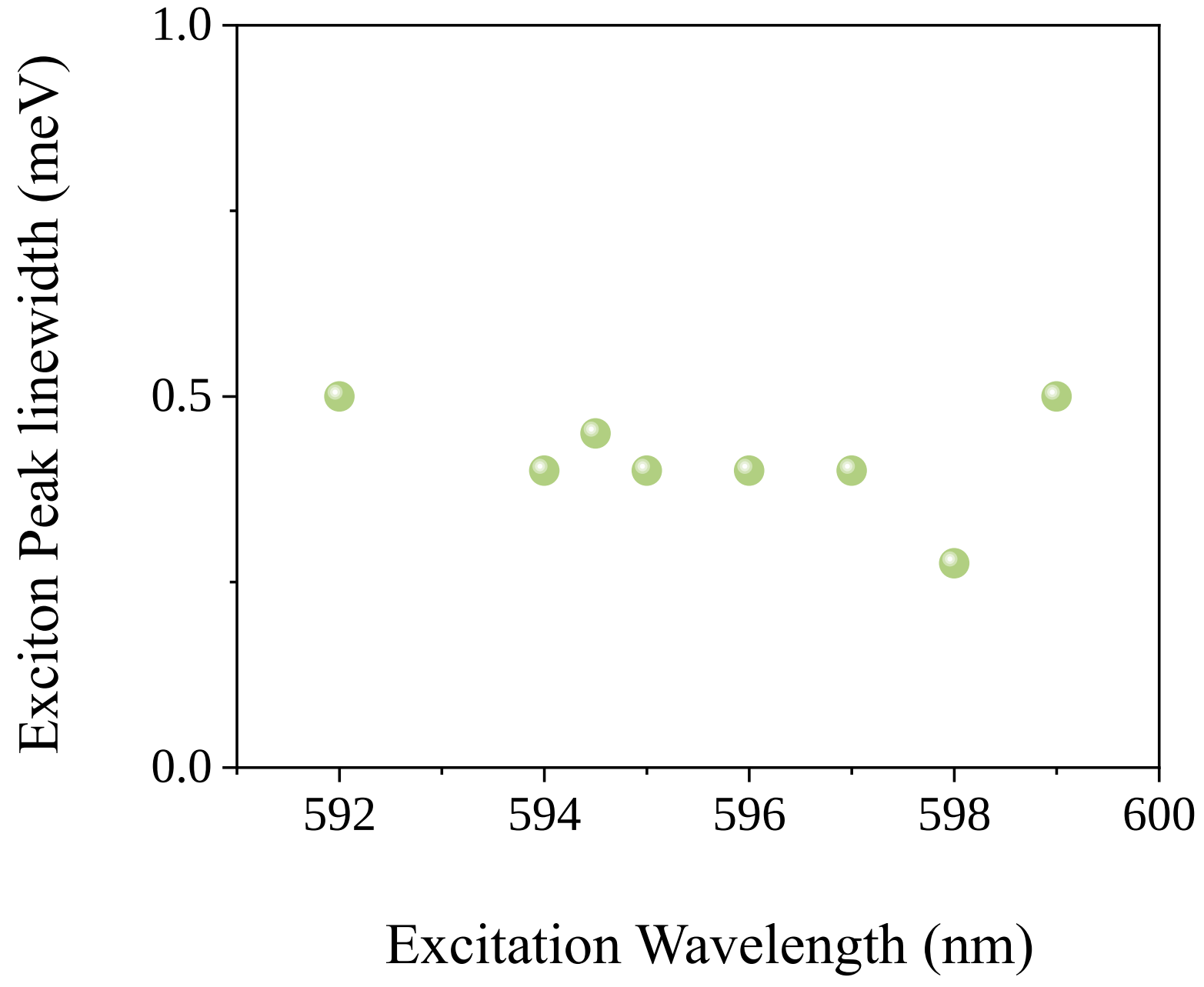

**Fig. S5:** Result from a different few-layer graphene capped monolayer $WS_2$ sample

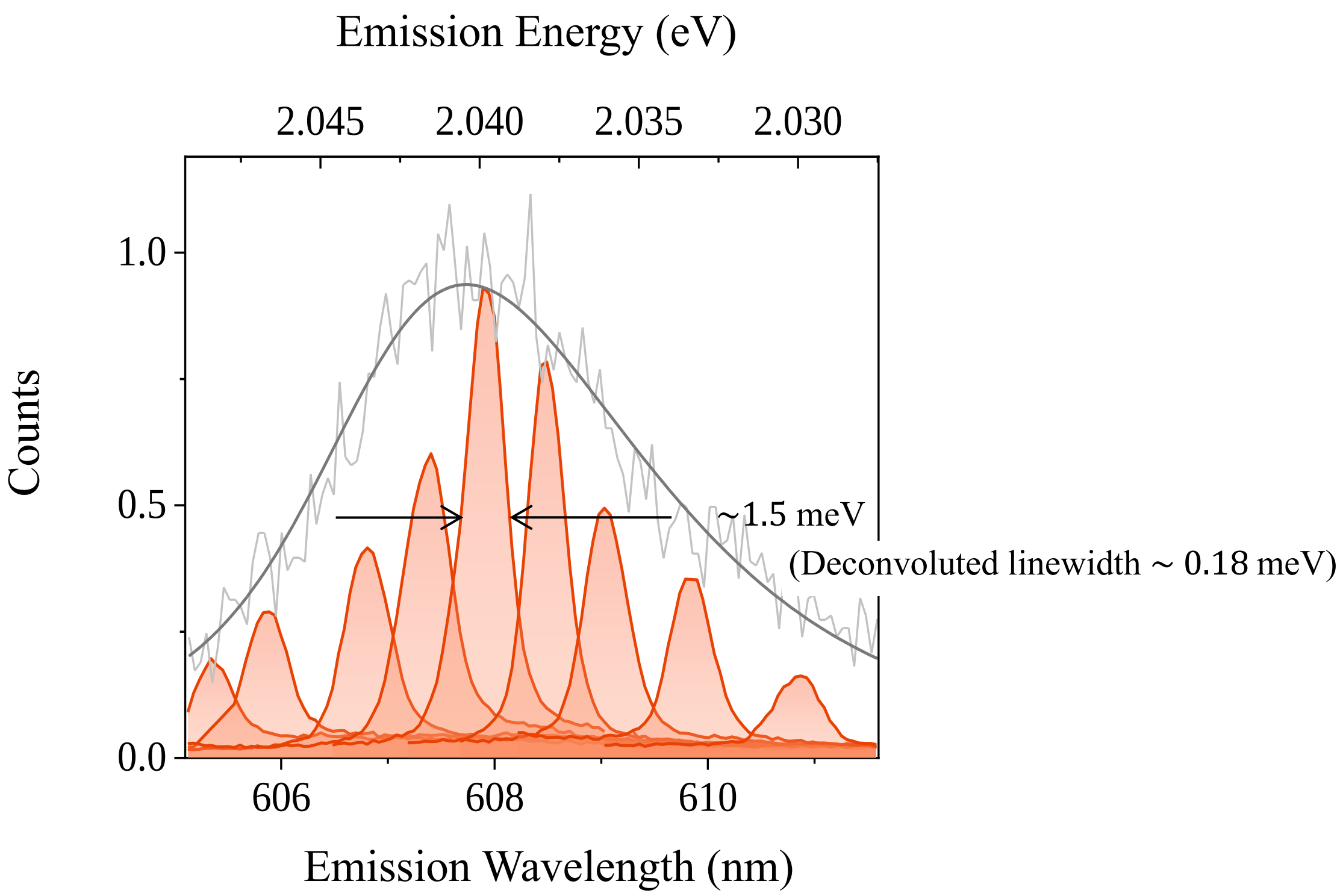

**Fig. S6:** Spectrum (in symbols) of the 1*s* exciton emission corresponding to the lowest linewidth obtained with 595 nm excitation, corresponding to single step relaxation with the $A_1'$ phonon. Deconvoluting the linewidth with the excitation laser (shown in yellow shading) gives a homogeneous broadening of 0.2 meV (fitted result shown in line).

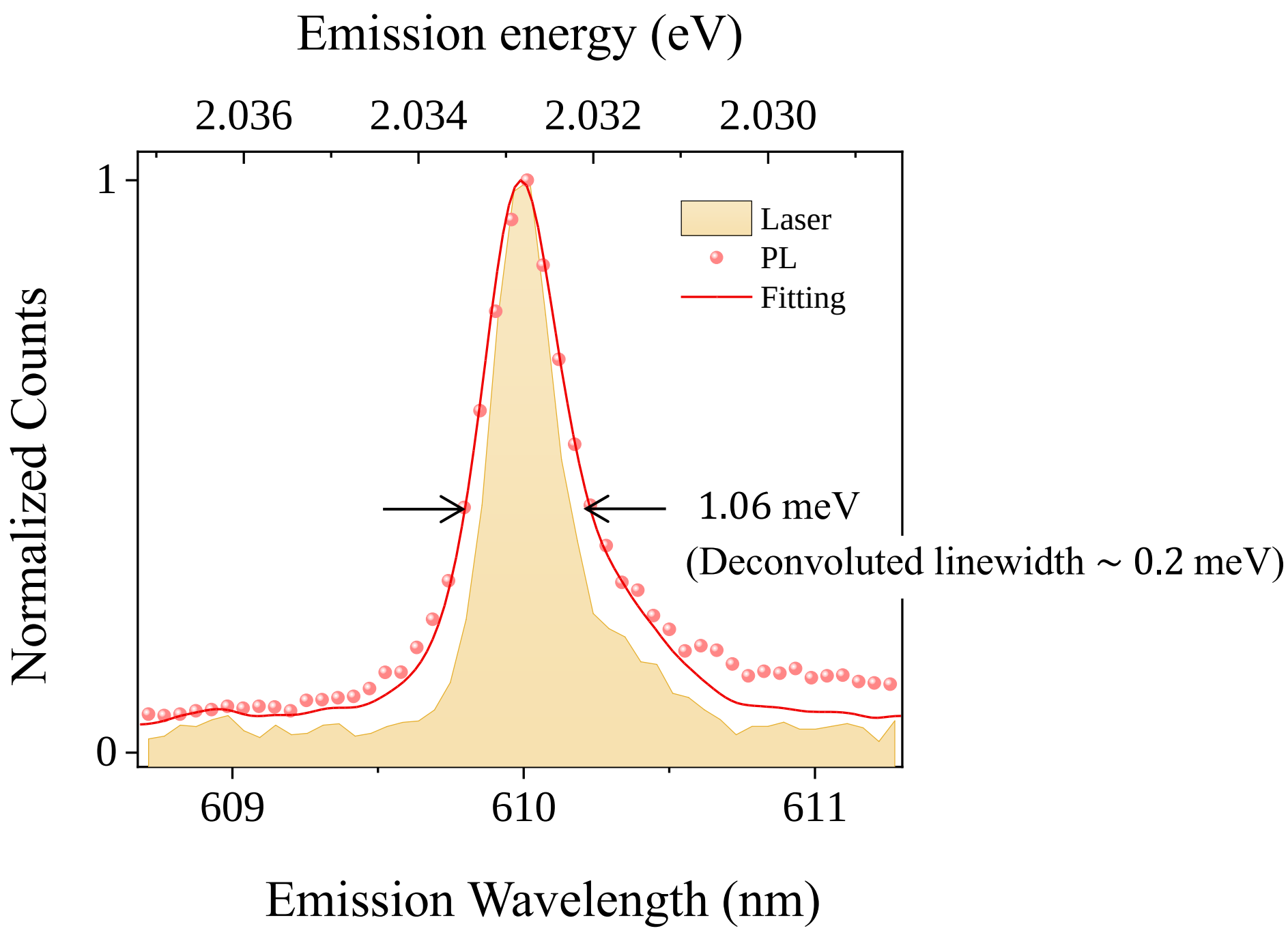

**Fig. S7:** Measurement taken on the GWG sample after two months of stack preparation.

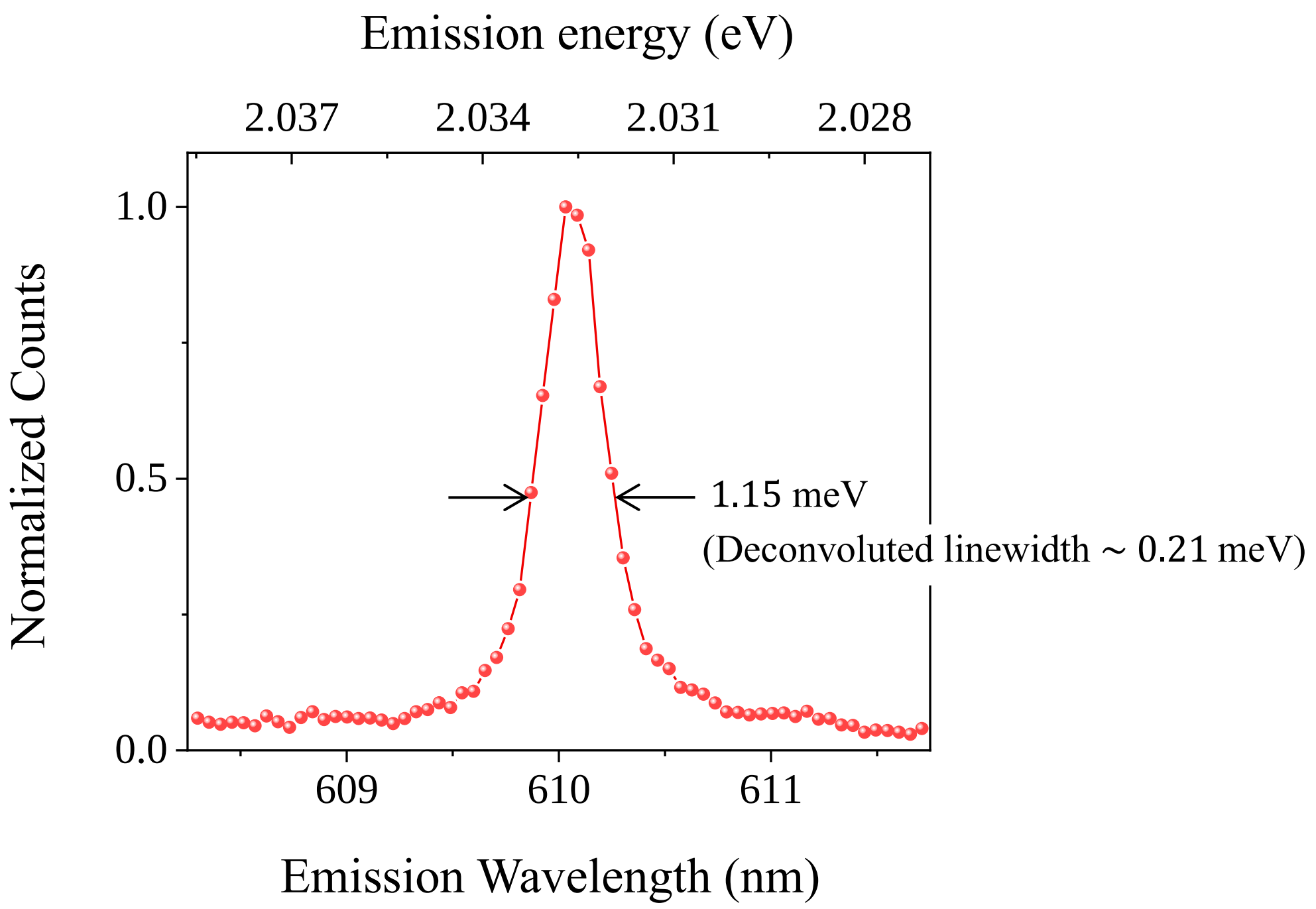

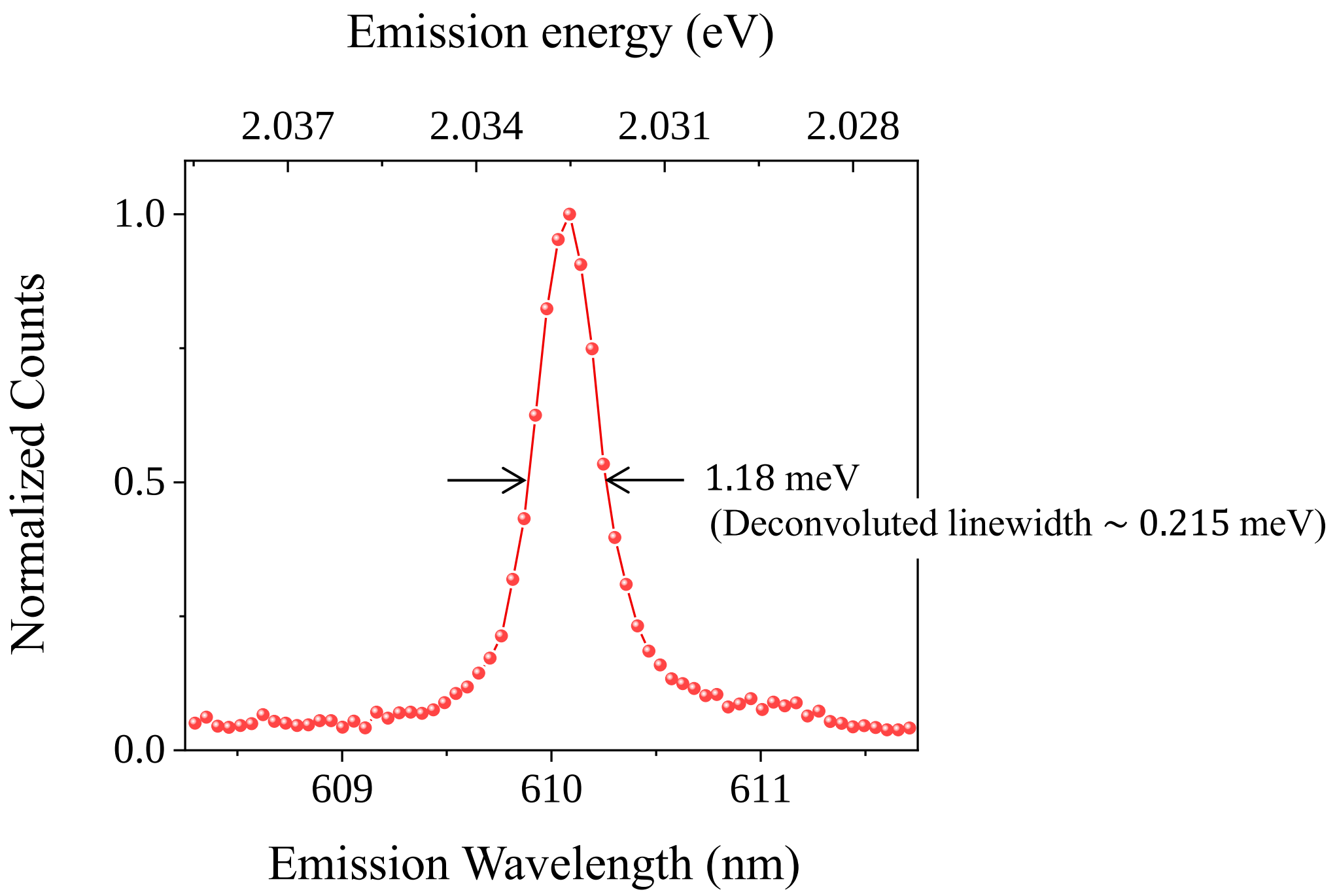